\documentclass[12pt,letterpaper]{article}
\usepackage{epsfig,rotating,setspace,latexsym,amsmath,epsf,amssymb,amsfonts,bm,theorem,cite,caption,subcaption,enumerate,longtable,accents,algorithm,graphicx,epsf,authblk,epstopdf,url,color,multirow,algpseudocode,mathtools,comment,tabularx,comment,physics,graphicx,caption,subcaption,braket,enumitem,xcolor,algorithm,braket}
\usepackage[short,c2]{optidef}
\usepackage[dvipsnames]{xcolor}
\usepackage[toc,title]{appendix}
\usepackage[colorlinks=true, linkcolor=blue]{hyperref}

\newtheorem{theorem}{Theorem}

\newtheorem{corollary}{Corollary}
\newtheorem{definition}{Definition}
\newtheorem{remark}{Remark}
\newtheorem{lemma}{Lemma}

\newenvironment{Proof}[1]{\medskip\par\noindent{\bf Proof:\,}\,#1}{{\mbox{\,$\blacksquare$}\par}}

\newcommand{\blue}[1]{{\color{blue}#1}}

\newcommand{\defeq}{\stackrel{\Delta}{=}}
\newcommand{\olX}{\overline X}
\newcommand{\olZ}{\overline Z}
\newcommand{\CX}{\operatorname{CX}}
\newcommand{\CZ}{\operatorname{CZ}}

\usepackage{array}
\usepackage{arydshln}
\usepackage{multirow}

\allowdisplaybreaks

\title{Entanglement Distribution in Quantum Switch Networks: Memory Limits and Achievability}

\author{Lei Hu \quad Alptug Aytekin \quad Stavros Mitrolaris \quad Mohamed Nomeir \quad Subhankar Banerjee   \quad Sennur Ulukus \quad Saikat Guha\\
\normalsize Department of Electrical and Computer Engineering\\
\normalsize University of Maryland, College Park, MD 20742 \\
\normalsize \emph{leihu@umd.edu} \quad \emph{aaytekin@umd.edu}  \quad \emph{stavros@umd.edu} \quad   \emph{mnomeir@umd.edu} \quad \emph{sbanerje@umd.edu}  \quad \emph{ulukus@umd.edu} \quad \emph{saikat@umd.edu}}

\begin{document}
\date{}
\maketitle
\vspace*{-0.88cm}
\begin{abstract}
    We study the minimum persistent memory required to distribute a graph state through a quantum switch. Specifically, consider a central switch that assists $n$ remote users in distributing a graph state, using independently generated user-switch Bell pairs. Since Bell-pair generation is probabilistic, different links generally succeed at different times. When a Bell pair is generated, the qubits may need to be stored in memory while waiting for other links to succeed. Recently, new protocols have been proposed to reduce this storage overhead by processing and releasing some switch qubits before all Bell links are available. However, these protocols do not characterize the minimum persistent switch memory required at each stage. Moreover, the local-cover method in the state-of-the-art Piecemaker protocol is computationally expensive for general target graphs. 
    
    In this work, we characterize the minimum persistent switch memory required after any subset of users' Bell links are generated and processed. We then develop an online stabilizer-transition protocol that attains this limit, which can be implemented in polynomial time in the number of qubits $n$ and does not require information about future Bell-link successes. For any graph and arrival order, our protocol requires less or equal persistent memory than the state-of-the-art Piecemaker at any stage of the distribution process. Moreover, when one Bell link succeeds per round of duration $\Delta t$, for the path graph $P_n$, Piecemaker requires the cumulative switch-qubit storage time $\Theta(n^2\Delta t)$ and peak memory of at least $\left\lfloor n/2 \right\rfloor$ \textit{for any arrival order}, while under some arrival orders, our protocol requires only $(n-1)\Delta t$ and retains just one persistent switch qubit. More generally, for tree graphs, our protocol can reduce the $\Theta(n^2\Delta t)$ cumulative storage time in Piecemaker to $O(n\log n\,\Delta t)$ for some arrival orders. Finally, the same stabilizer-transition framework also provides a direct forward construction for sequential graph-state generation using the minimum number of quantum emitters, equal to the maximum of the height function.
\end{abstract}

\section{Introduction}
Entanglement is a fundamental resource for quantum information processing and quantum networks \cite{shor1995scheme, gottesman2024surviving, kitaev2003fault}. In multipartite settings, a particularly important class of entangled states is stabilizer states, which can be efficiently described and manipulated using Clifford operations \cite{dehaene2003clifford}. Every stabilizer state is local-Clifford (LC) equivalent to a graph state, which provides a convenient graphical representation of its entanglement structure. Graph states are key resources for quantum computation and distributed quantum protocols \cite{hein2006entanglement,raussendorf2001one,sen2025multipartite}. However, realizing these applications over a quantum network requires graph states to be distributed among remote nodes (i.e., users). This distribution is challenging, as noise introduced during transmission, storage, and processing can degrade the fidelity of the resulting state.

\begin{figure}[t] 
    \centering
    \includegraphics[width=0.7\textwidth]{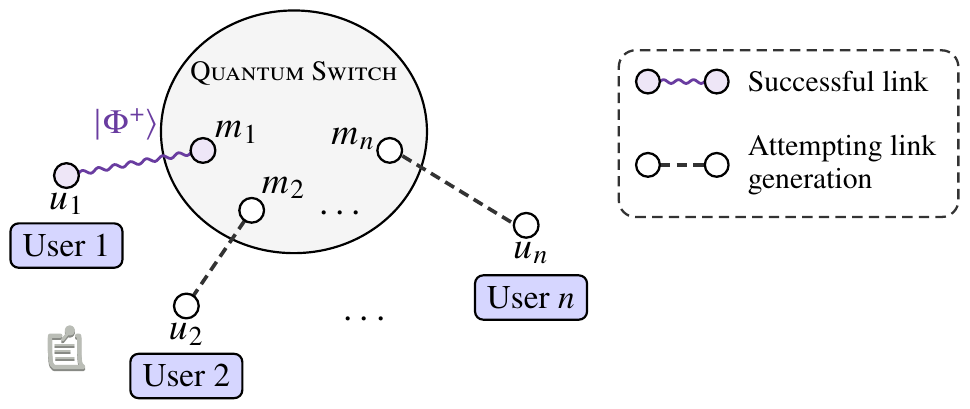}
    \caption{System model of the quantum switch network. Each user is connected to the central switch through a quantum channel. When Bell-pair generation on a user-switch link succeeds, the two halves $(u_i,m_i)$ are stored at the user and the switch, respectively. Otherwise, the corresponding memory remains empty.}
    \label{fig:switch_model}
\end{figure}

Recently, the concept of a quantum switch has been proposed as a scheme for entanglement distribution among remote users \cite{mitrolaris2026age,fischer2021distributing,vardoyan2019stochastic,promponas2023full,vardoyan2023capacity,dai2022capacity,banerjee2026index}. Specifically, as illustrated in Fig.~\ref{fig:switch_model}, consider a central quantum switch connected to $n$ users through independent quantum channels. Each user and the switch first attempt to establish a Bell pair between the user's memory subsystem, represented by $u_i$ and the switch's memory subsystem $m_i$. Once the Bell pairs are available, the switch performs local operations to distribute the target graph state among the users. However, the generation of Bell links is probabilistic and thus different links can succeed at different times. A conventional scheme, referred to as the \textit{Factory} protocol \cite{avis2023analysis}, waits until all Bell pairs are available. It then prepares the target state at the switch and teleports its qubits to the users' qubits. Consequently, when a Bell pair is generated early, its switch-side qubit has to stay in memory until the remaining links are established. The user-side qubit also needs to be stored since it will be part of the final graph state. However, the switch-side qubit is only needed during the distribution process. While a switch qubit waits, memory errors can reduce the fidelity of its shared Bell pair \cite{li2021efficient,davies2024entanglement,hermans2022qubit}. Teleportation through that noisy pair can lower the fidelity of the final graph state \cite{cuquet2012growth}. Also, the occupied switch memory will be unavailable for other tasks.

More recently, \textit{Piecemaker} protocol \cite{prielinger2026piecemaker} has been introduced to reduce this storage overhead by waiting only until the available Bell links contain a minimal local cover (MLC) of the target graph, i.e., a vertex cover of some graph that is LC-equivalent to the target. Then, subsequent Bell-pair halves can be processed and released once their links succeed. However, Piecemaker requires computationally expensive preprocessing to identify MLCs over graphs that are LC-equivalent to the target graph \cite{sharma2026minimising}.
In this paper, we consider going beyond this local cover approach. Specifically, we investigate the minimum number of \textit{persistent} qubits that the switch must retain over the distribution process. Here, persistent qubits refer to the switch qubits that remain stored after a processing step and are retained while waiting for future Bell-link successes. After any set of users has been processed, we aim to characterize the minimum number of persistent qubits required at the switch. This minimum gives a fundamental memory limit for graph state distribution. However, achieving this limit is nontrivial. First, the processed users must already have the same reduced state, as the switch can no longer directly apply joint quantum operations on these users' qubits after they have been processed, i.e., only local operations can be performed on these qubits. Second, the quantum correlations between the processed and unprocessed users must still be preserved through the switch. Finally, as future Bell links arrive in an unknown order, the switch must maintain and update these correlations online while retaining as few qubits as possible.

Inspired by recent works on sequential photonic graph-state generation \cite{schoen2005sequential, emitter2022,kaur2025resource}, we connect the switch-memory requirement to the resources required for sequential graph-state generation. In particular, Li \textit{et al.}~\cite{emitter2022} considered sequential graph state generation using the minimum number of quantum emitters. Quantum emitters are stationary quantum systems that generate and interact with photonic qubits. After a photonic qubit is generated and interacts with the emitters, it is emitted and no longer participates in subsequent operations. The emitter retains the quantum correlations between the emitted photons and those that have not been generated. For a fixed order in which the photonic qubits are generated, \cite{emitter2022} showed that the required number of emitters is determined by the maximum bipartite entanglement across the emitted and unemitted cut. They further developed a time-reversed algorithm that starts from the target graph state and sequentially absorbs the photonic qubits into the emitters. Then, reversing this procedure yields the sequence of emitter operations and photon emissions for generating the target state. This provides a natural connection to the quantum-switch setting.

Despite this connection, their model differs from the quantum-switch setting in two important aspects. First, in the quantum-switch model, Bell links are established probabilistically, so the order in which users can be processed is generally not known in advance. In contrast, the photon emission order in sequential generation is prescribed before the generation process. Second, in sequential generation, an emitter locally generates each photonic qubit and can directly interact with it. However, in the quantum-switch setting, the user qubits are located remotely from the switch, so the switch cannot directly apply local operations on them. When a link succeeds, the switch only holds one half of the Bell pair, i.e., $m_i$ in Fig.~\ref{fig:switch_model}, while the other half, $u_i$, is located at the remote user side. Therefore, the switch cannot directly perform the same joint operation between its memory qubit and the user qubit as in sequential generation. Hence, the sequential state generation protocol cannot be applied to the quantum switch setting.

Our contributions in this paper are as follows.
\begin{itemize}
    \item We establish a fundamental lower bound on the persistent switch memory required after any subset of users' Bell links are generated and processed. In particular, for an arbitrary pure target state, the bound is the base-two logarithm of the rank of the reduced density operator on the processed users. For stabilizer states, the bound equals the von Neumann entropy of the reduced state on these processed users. For graph states, it reduces to the binary cut-rank of the adjacency submatrix of processed-unprocessed users partition.

    \item We develop an online stabilizer-transition protocol that attains the memory lower bound after processing each newly established Bell link. The Bell-link arrival order is completely unknown to the protocol, and no statistical assumptions are made about future Bell-link arrivals. In particular, at each step, we maintain a \emph{minimal stabilizer purification }of the target reduced state on the processed users. This is a stabilizer state shared by the processed users and the switch. It is minimal because the switch retains the minimal number of persistent qubits required by the lower bound.
    We give explicit stabilizer-transition operations that update this minimal purification whenever a new user is processed. Moreover, our protocol can be implemented in polynomial time and the worst-case classical complexity is $O(n^4)$.

    \item We compare the proposed protocol and the state-of-the-art heuristic Piecemaker protocol~\cite{prielinger2026piecemaker}. For  distributing the path graphs with $n$ users, Piecemaker requires at least $\lfloor n/2\rfloor$ peak persistent switch qubits for any arrival order, while our protocol requires just one switch qubit over the distribution process under some orders. We further compare the cumulative switch-qubit storage time when one Bell link succeeds per round of duration $\Delta t$. In particular, for path graphs $P_n$, Piecemaker requires $\Theta(n^2\Delta t)$ cumulative storage time for every arrival order, while our protocol achieves $(n-1)\Delta t$ cumulative storage time for a suitable order. More generally, for bounded-degree tree graphs, our protocol can achieve $O(n\log n\,\Delta t)$ cumulative storage time compared with $\Theta(n^2\Delta t)$ for Piecemaker and Factory protocols under some arrival orders.

    \item Finally, the stabilizer-transition framework provides a direct forward construction for sequential graph state generation using the minimum number of quantum emitters. Unlike the existing time-reversed construction in \cite{emitter2022}, our protocol proceeds directly along the generation order by maintaining and updating a minimal stabilizer purification after each emitted qubit. The number of required emitters then is equal to the maximum of the corresponding {\em height function} of the target graph state~\cite{emitter2022}.
\end{itemize}

\textit{Notation:}
$\mathbb{F}_{2}$ denotes the finite field of order $2$, and $\mathbb{F}_{2}^{a\times b}$ denotes the set of $a\times b$ matrices over $\mathbb{F}_{2}$. The symbol $\defeq$ means ``defined as''. For a set $\mathcal{I}$, $\mathcal{I}\setminus\{i\}$ denotes the set obtained by removing $i$ from $\mathcal{I}$, while $\mathcal{I}\setminus\mathcal{J}$ denotes the set obtained by removing all elements of $\mathcal{J}$ from $\mathcal{I}$. For a stabilizer state $\ket{\psi}$, $S(\ket{\psi})$ denotes its stabilizer group. For a Pauli operator $P$, $\mathrm{supp}(P)$ denotes the set of qubits on which $P$ acts nontrivially. For a stabilizer group $S$, $N(S)$ denotes its Pauli normalizer, i.e., the set of Pauli operators that commute with every element of $S$. The quotient group $N(S)/S$ represents the logical Pauli operators associated with $S$. We use $\rank_{\mathbb C}(\rho)$ for the matrix rank of a density operator $\rho$ over $\mathbb C$, and $\rank_{\mathbb{F}_2}(M)$ for the rank of a binary matrix $M$ over $\mathbb{F}_2$. For $x\in \mathbb{R}$, $\lfloor x\rfloor$ denotes the greatest integer not larger than $x$, and $\lceil x\rceil$ denotes the smallest integer not smaller than $x$. For a set of operators $S$, $\langle S\rangle$ denotes the group generated by the elements of $S$. Moreover, $\langle S, g_1,\ldots,g_k\rangle$ denotes the group generated by $S$ and $g_1,\ldots,g_k$.

\section{Problem Formulation}
\subsection{Quantum Network Model}
We consider a star-topology network consisting of a central quantum switch and a set of remote users indexed by $V \defeq \{1,2,\ldots,n\}$.
The switch is equipped with $n$ quantum memory qubits $\{m_1,\ldots,m_n\}$ and assists the users in distributing a prescribed stabilizer state.
Specifically, each user $i\in V$ has a single quantum memory qubit $u_i$ and is connected to the switch through a quantum channel. The goal of the switch is to distribute the stabilizer state among the remote user qubits $u_i$, $i\in V$. Since any pure stabilizer state is local-Clifford equivalent to a graph state, in this work, we focus on distributing graph states, and the same framework also applies to pure stabilizer states up to local Clifford operations.

\begin{definition}(Graph State)
Let $G = (V,E)$ be an undirected graph. The associated $n$-qubit graph state is
\begin{align}
    \ket{G}_V \defeq \prod_{\{i,j\}\in E}\CZ_{ij}\ket+^{\otimes n}.
\end{align}
\end{definition}
Equivalently, $\ket{G}_V$ is the unique simultaneous $+1$ eigenstate of the stabilizer generators
\begin{align}
    K_i=X_i\prod_{w\in N_G(i)}Z_w, \quad i\in V,
\end{align}
where $N_G(i)$ denotes the neighborhood of vertex $i$ in $G$.

The target graph $G$ is fixed and known to the switch before the distribution protocol begins. To distribute $\ket{G}$ among the users, Bell pairs are first generated over the individual user-switch links, with $\ket{\Phi^+}_{u_i m_i}$ shared between user $i$'s memory $u_i$ and the switch's memory $m_i$. The switch then processes the successfully generated Bell pairs using local quantum operations. More specifically, the distribution process consists of the following two interleaved stages.

\paragraph{Stage I: Bell-link generation.}
To establish entanglement between $u_i$ and $m_i$, the corresponding quantum link repeatedly attempts to generate a Bell pair until success. Upon a successful attempt, $u_i$ and $m_i$ share the Bell state
\begin{align}
    \ket{\Phi^+}_{u_i m_i} = \frac{\ket{00}_{u_i m_i} + \ket{11}_{u_i m_i}}{\sqrt{2}}.
\end{align}
Each attempt is heralded, so both the user $i$ and the switch know whether it succeeded. In addition, different quantum links operate independently in the same manner. Following the network model of Piecemaker \cite{prielinger2026piecemaker}, entanglement generation proceeds in discrete rounds of duration $\Delta t$. If multiple links succeed in the same round, we treat them as one batch, and the switch processes the batch in any preferred internal order. Note that the Bell links generally become available at different times, and their order is completely unknown.

\paragraph{Stage II: Switch processing.}
Whenever a Bell link is established, the switch can process the new switch-side qubit $m_i$ and the retained qubits. To process the qubits, the switch is equipped with a quantum processor that can perform controlled-$Z$ (CZ) and controlled-$X$ (CX) gates between its memory qubits, single-qubit Clifford gates, and single-qubit Pauli measurements.\footnote{In fact, only one of CZ and CX is required, because they are equivalent up to single-qubit Hadamard gates.} It can also communicate heralding signals and measurement outcomes classically to the users. The remote users are restricted to single-qubit Clifford gates and Pauli measurements. 

Following the assumption in \cite{prielinger2026piecemaker}, we further assume that local quantum operations, measurements, and classical feed-forward are sufficiently fast compared with the entanglement-generation time $\Delta t$. After processing, a switch-side qubit may either be released or retained for future Bell-link successes. We refer to the qubits that remain stored at the switch after each processing step as the \emph{persistent switch memory}. Note that in physical implementations, stored qubits can decohere while waiting for future links. The resulting errors can degrade the shared entanglement and lower the fidelity of the final graph state. Moreover, releasing switch qubits also makes their quantum memory available for other tasks. Therefore, our goal is to minimize this persistent memory while eventually distributing the target graph state $\ket{G}_V$. 

\subsection{Existing Protocols}
We next review two existing approaches to graph-state distribution under the network model above. We illustrate both protocols using the four-qubit path $P_4:1-2-3-4$ and the same Bell-link arrival times. We will return to this path graph when illustrating the proposed protocol in Section~\ref{sec:example}.

\begin{figure}[!t]
    \centering
    \includegraphics[
        width=\linewidth
    ]{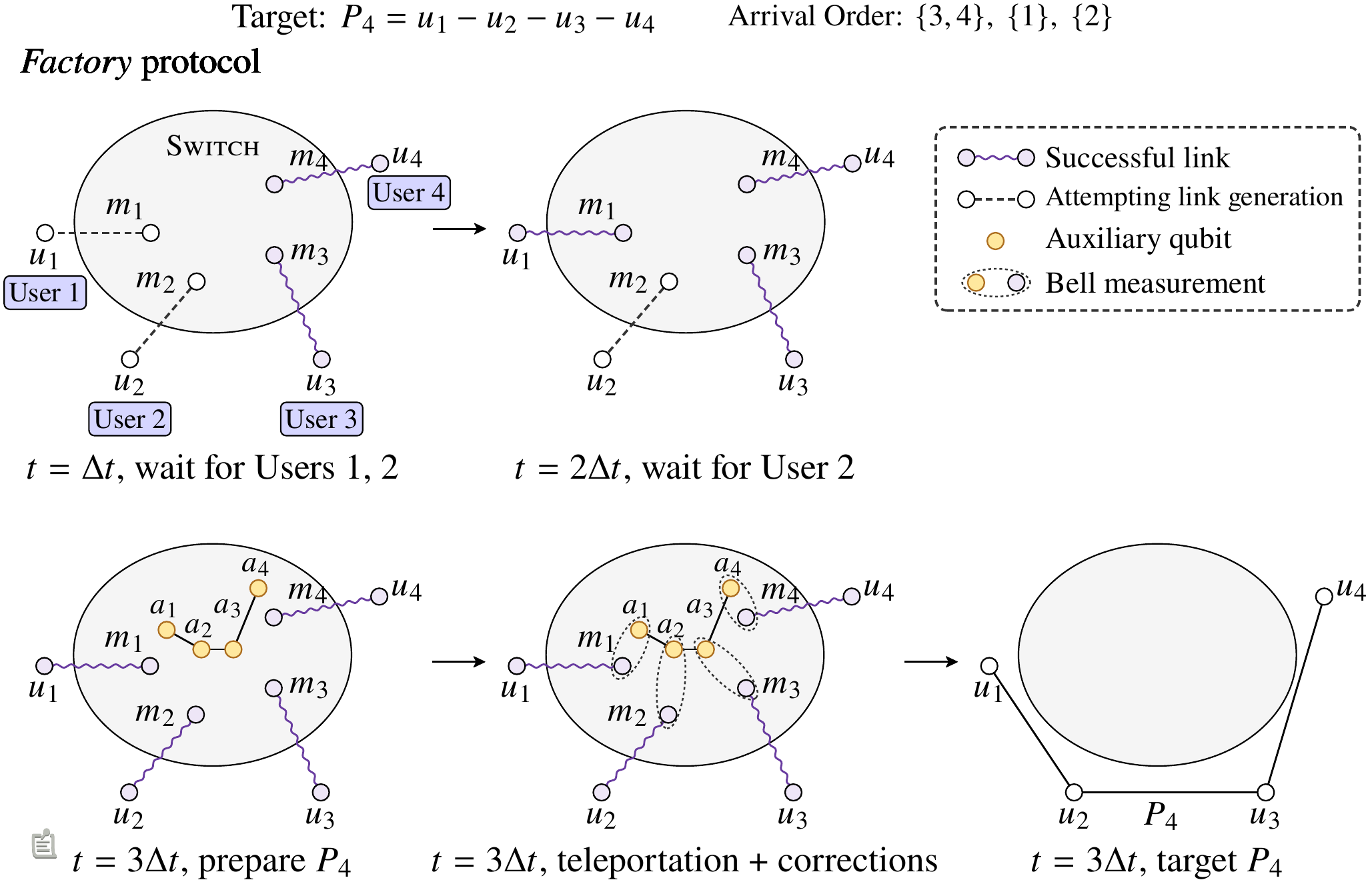}
    \caption{Factory protocol for distributing the four-qubit path graph state. Users $3$ and $4$ establish Bell links at $t = \Delta t$, while users $1$ and $2$ generate the links at $t = 2 \Delta t$ and $t = 3 \Delta t$, respectively. The switch stores its halves of the successful Bell pairs until all four links are available. It then prepares the target state on four auxiliary qubits $a_1,\ldots,a_4$.
    The switch performs a Bell-state measurement on each pair $(a_i,m_i)$ to teleport the state of $a_1,\ldots,a_4$ to the user qubits $u_1,\ldots,u_4$. Each user applies a single-qubit Pauli correction based on the corresponding measurement outcome. The users then share the target graph state.}
    \label{fig:factory}
\end{figure}

\begin{figure}[!t]
    \centering
    \includegraphics[width=0.99\textwidth]{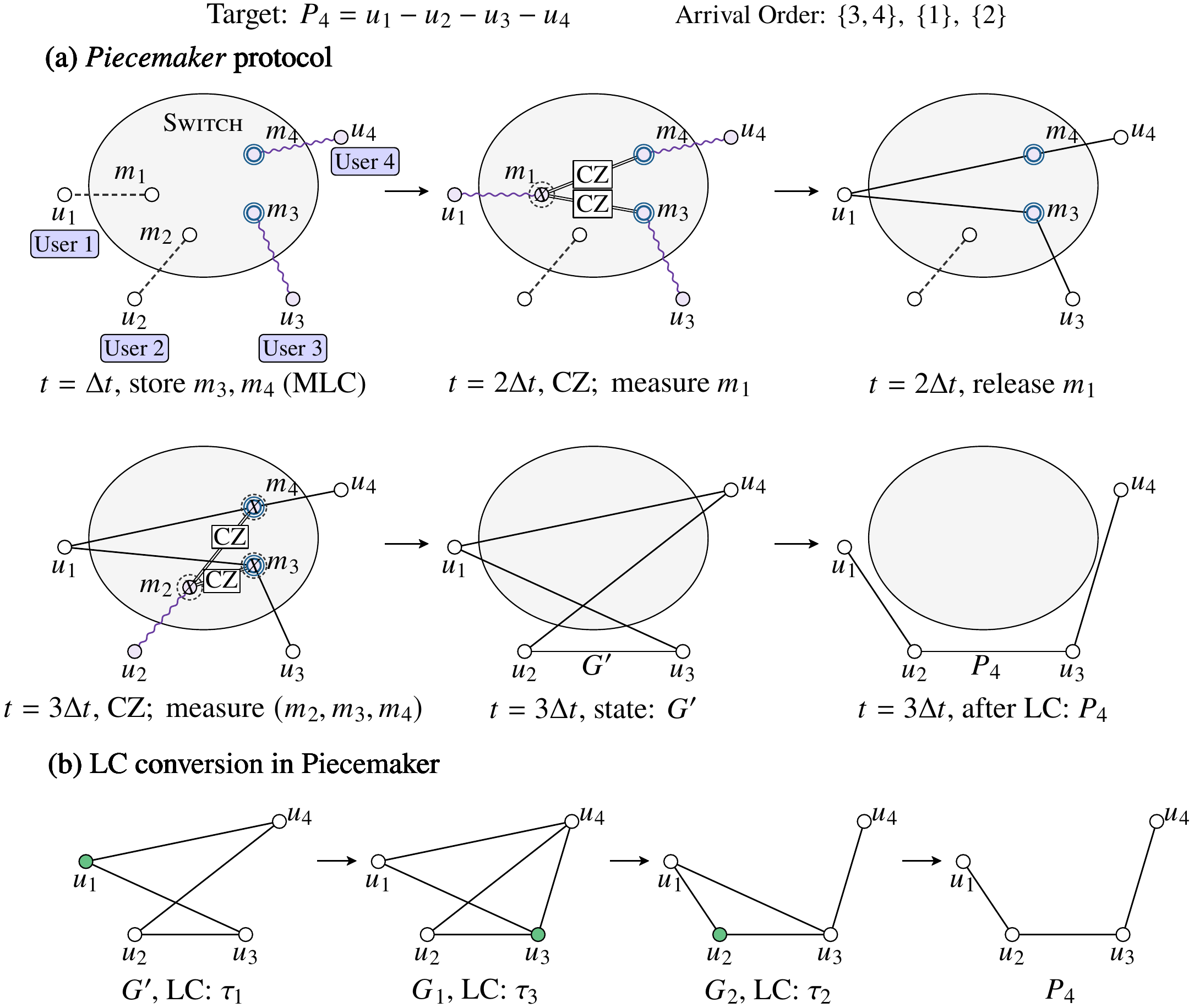}
    \caption{Piecemaker protocol for distributing the four-qubit path graph state, where users $3$ and $4$ establish their Bell links at $t = \Delta t$, while users $1$ and $2$ succeed at $t = 2 \Delta t$ and $t = 3 \Delta t$, respectively.
    (a) When users $3$ and $4$ generate their Bell links, the set $C=\{3,4\}$ is a vertex cover of the LC-equivalent graph $G'$. The switch then retains $m_3$ and $m_4$ as persistent qubits, and processes each later link upon arrival. It releases $m_1$ at $t=2\Delta t$ and all remaining switch qubits at $t=3\Delta t$. After all Bell links are generated and processed, the users' qubits $u_1,\ldots,u_4$ form the graph state $G'$. By performing the single-qubit LC operations $\tau_1,\tau_3,\tau_2$, the users share the target graph state $P_4$.
    (b) Local complementations $\tau_1, \tau_3,\tau_2$ applied at $u_1$, $u_3$, and $u_2$, respectively, transform $G'$ into $P_4$. Each operation $\tau_v$ adds missing edges and removes existing edges between the neighbors of $v$. For example, the neighbors of $u_1$ in $G'$ are $u_3$ and $u_4$. Since they are not connected by an edge, $\tau_1$ adds the edge $u_3-u_4$, which produces the graph $G_1$.} \label{fig:factory_piecemaker}
\end{figure}
\emph{Factory Protocol:}
The conventional approach, referred to as the \emph{Factory} protocol \cite{avis2023analysis}, waits until all Bell links have been established. The switch then locally prepares the target graph state $\ket{G}$ on $n$ auxiliary memory qubits and teleports their joint state to the corresponding remote users using the established Bell pairs. As a result, the switch halves of Bell pairs established early must remain stored until all links become available. This increases their exposure to memory decoherence and can reduce the fidelity of the distributed graph state. 

Figure~\ref{fig:factory} illustrates this procedure for $P_4$. At $t = \Delta t$, the switch holds $m_3, m_4$, while at $t=2\Delta t$, it holds $m_1$, $m_3$, and $m_4$. It must retain these qubits until the link for user~$2$ succeeds at $t=3\Delta t$. Then, it prepares and teleports the target state.

\emph{Piecemaker Protocol:}
Piecemaker~\cite{prielinger2026piecemaker} reduces this storage overhead by processing some Bell links before all links are available. In particular, it waits until the available Bell links contain a minimal local cover (MLC) of the graph. First, recall that a subset $W\subseteq V$ is a \emph{vertex cover} of $G$ if every edge in $E$ has at least one endpoint in $W$. A vertex cover $W$ is \emph{minimal} if no strict subset of $W$ is also a vertex cover. Now, suppose that the established Bell links contain such a vertex cover $W$. For any subsequently established link corresponding to $v\in V\setminus W$, all neighbors of $v$ belong to $W$. Then, the switch measures the corresponding stabilizer $K_v=X_{u_v}\prod_{w\in N_G(v)}Z_{u_w}$, where $N_G(v)$ denotes the neighboring vertices to $v$ in the graph $G$, which depends only on $v$ and vertices whose links are already available. Piecemaker can therefore impose this stabilizer and release the corresponding switch-side qubit. Thus, the protocol does not need to wait for the remaining links. 

Furthermore, Piecemaker exploits local-Clifford equivalence. In particular, the switch may use a graph $G'$ that is local-Clifford equivalent to the target graph $G$. If the available Bell links contain a vertex cover $C$ of $G'$, newly established links outside the cover can again be processed immediately. This motivates the following definition.  
\begin{definition}
    A set $C \subseteq V$ is a local cover of $G$ if it is a vertex cover of some graph $G'$, where $G'$ is local-Clifford equivalent to $G$. It is a minimal local cover (MLC) if no proper subset is also a local cover of $G$.
\end{definition}
After the distribution of $\ket{G'}$ is completed, local Clifford operations transform it into the target state $\ket{G}$.

Figure~\ref{fig:factory_piecemaker} illustrates this approach for the same path graph and arrival times. The set $C = \{3,4\}$ is not a vertex cover of graph $P_4$, but it is a vertex cover of the LC-equivalent graph $G'$ shown in the figure. Thus, a local cover is already available at $t = \Delta t$. When the Bell pair $\ket{\Phi^+}_{m_1 u_1}$ for user $1$ succeeds at $t = 2 \Delta t$, the switch processes and releases $m_1$. Only $m_3$ and $m_4$ are stored while waiting for user $2$, whereas Factory protocol retains three switch qubits during the same arrival. After all links have been processed, the users share the graph state $\ket{G'}$. Then, the users perform sequential local complementations shown in Figure~\ref{fig:factory_piecemaker}(b) to convert the graph $G'$ to the target $P_4$.

As a result, Piecemaker can release some switch qubits before all Bell links are available, thereby reducing the persistent switch memory. However, the algorithm to find the MLCs requires substantial computational complexity \cite{sharma2026minimising}. This limits its use for larger graphs. Therefore, our motivation in this paper comes from the two questions not answered by the Piecemaker: what is the \textit{minimum number of persistent switch qubits} and how to \textit{achieve it with polynomial classical computation}. To this end, we first derive a memory lower bound for arbitrary pure states. We then show that this bound is achievable for graph states.

\section{Minimum Memory at the Switch} \label{sec:memory_limit}
Let $A\subseteq V$ and $\bar{A} \defeq V \setminus A$ denote the set of processed users and unprocessed users, respectively. We first establish a lower bound on the required switch memory. This bound applies to any general pure state $\ket{\psi}_V$ to be distributed among the users $V$.

\subsection{Lower Bound for General Pure States}
Let $E_A$ denote the quantum register retained at the switch side after processing $A$, and let $q_A \defeq |E_A|$ be the number of qubits of $E_A$. $\rho_A \defeq \Tr_{\bar{A}} \ketbra{\psi}$ is the reduced state of the target state on the processed users. For any valid protocol, the intermediate state need not be pure. To derive a memory bound that also applies to mixed intermediate states, we use the Schmidt number, which extends Schmidt rank to mixed states. 
For completeness, we first recall its definition \cite{terhal2000schmidt,nielsen2010quantum}.
\begin{definition}
    A bipartite density operator $\rho_{XY} $ on $ \mathcal{H}_X \otimes \mathcal{H}_Y$ has Schmidt number $k$ across the bipartition $X|Y$, denoted as $\mathrm{SN}_{X|Y}(\rho_{XY}) = k$, if:
    \begin{enumerate}
        \item For any pure-state decomposition of $\rho_{XY}$, $\{p_i> 0, \ket{\psi_i}\}$, such that 
        \begin{align}
            \rho_{XY} = \sum_i p_i \ketbra{\psi_i}_{XY},
        \end{align}
        at least one of $\{\ket{\psi_i}_{XY}\}$ has Schmidt rank at least $k$, i.e., $\max_i \mathrm{SR}_{X|Y}(\ket{\psi_i}_{XY}) \geq k$. Here, $\mathrm{SR}_{X|Y}(\ket{\psi_i}_{XY})$ denotes the Schmidt rank of $\ket{\psi_i}$ across $X|Y$. It is the number of nonzero coefficients in its Schmidt decomposition.

        \item There exists a decomposition of $\rho_{XY}$ in which all $\{\ket{\psi_i}_{XY}\}$ have Schmidt rank at most $k$, i.e., $\max_i \mathrm{SR}_{X|Y}(\ket{\psi_i}_{XY}) \leq k$.
    \end{enumerate}
\end{definition}
Equivalently, we can write
\begin{align}
    \mathrm{SN}_{X|Y}(\rho_{XY}) = \min_{ \rho_{XY} = \sum_i p_i\ketbra{\psi_i}_{XY}} \max_i \ \mathrm{SR}_{X|Y}(\ket{\psi_i}_{XY}).\label{eq:def_SN}
\end{align}
For a pure state, the Schmidt number is its Schmidt rank \cite{terhal2000schmidt}. Moreover, the Schmidt number is nonincreasing under local operations and classical communication (LOCC).

We now bound the persistent memory needed to distribute a pure state.
\begin{theorem}\label{thm:memory_lower_bound}
    Consider any exact protocol that distributes a pure state $\ket{\psi}_V$ under the network model above. After processing any subset of users $A \subseteq V$, the number of persistent switch qubits satisfies
    \begin{align}
        q_A \geq \left\lceil \log_2 \rank_{\mathbb C} \left( \rho_A \right) \right\rceil .\label{eq:general_memory_bound}
    \end{align}
\end{theorem}

\begin{Proof}
    Consider the Schmidt decomposition of the target state across $A|\bar{A}$ \cite{peres1997quantum},
    \begin{align}
        \ket{\psi}_{A \bar{A}}=
        \sum_{j=1}^{R_A} \sqrt{\lambda_j} \ket{a_j}_A \ket{b_j}_{\bar{A}},
    \end{align}
    where $\{\ket{a_1},\ldots,\ket{a_{R_A}}\}$ and $\{\ket{b_1},\ldots,\ket{b_{R_A}}\}$ are orthonormal sets. The coefficients $\sqrt{\lambda_j} > 0$, $j=1,\ldots,R_A,$ satisfy $\sum_{j=1}^{R_A} \lambda_j = 1$. Here, $R_A \defeq \mathrm{SR}_{A|\bar{A}}(\ket{\psi})$ is the Schmidt rank across the bipartition $A|\bar{A}$. Also, $ R_A = \rank_{\mathbb C} \left( \rho_A \right) = \rank_{\mathbb C} \left( \rho_{\bar{A}} \right)$. 
    
    Now, consider an arbitrary distribution protocol that distributes the target state $\ket{\psi}$ among users $V = A \cup \bar{A}$. Let $\mathsf{A}$ denote all quantum systems held by the processed users in $A$, and $\omega_{\mathsf{A}E_A}$ denote the current joint state with the retained switch register $E_A$. In general, $\omega_{\mathsf{A}E_A}$ need not be pure. Since $E_A$ contains $q_A$ qubits, we have $\mathrm{dim} \mathcal{H}_{E_A} = 2^{q_A}$. Therefore, every pure state on $\mathcal{H}_{\mathsf{A}} \otimes \mathcal{H}_{E_A}$ has Schmidt rank at most 
    \begin{align}
        \min \{\dim \mathcal{H}_{\mathsf{A}}, \mathrm{dim} \mathcal{H}_{E_A}\} \leq 2^{q_A}.
    \end{align}
    By the definition of Schmidt number, we have
    \begin{align}
        \mathrm{SN}_{\mathsf{A}|E_A}(\omega_{\mathsf{A}E_A}) \leq 2^{q_A}.
    \end{align}
    Now, we fix a bipartition of the network nodes. The left side contains the users in $A$, while the right side contains the switch and all users in $\bar{A}$. We keep this node bipartition fixed throughout the rest of the proof. On the right side, we also explicitly include the Bell pairs that will be generated for users in $\bar{A}$. Specifically, let $T$ collect both halves of these pairs, and define
    \begin{align}
        \rho_{T} \defeq \bigotimes_{v \in \bar{A}} \ketbra{\Phi^+}_{u_v m_v},
    \end{align}
    where both $u_v$ and $m_v$ belong to the right side of the bipartition, as neither is a qubit of users in $A$.
    Appending $\rho_T$ is a right-side local operation and cannot increase the Schmidt number. Conversely, tracing out $T$ recovers $\omega_{\mathsf{A}E_A}$ and is also a right-side local operation. Hence, by the monotonicity of Schmidt number in both directions, we have
    \begin{align}
        \mathrm{SN}_{\mathsf{A} | E_A T} \left( \omega_{\mathsf{A}E_A} \otimes \rho_T \right) = \mathrm{SN}_{\mathsf{A} | E_A}\left( \omega_{\mathsf{A}  E_A} \right) \leq 2^{q_A}. \label{eq:tensor}
    \end{align}
    The equality in \eqref{eq:tensor} can also be proved using Schmidt decompositions. Specifically, let $\ket{\Phi}_T \defeq \bigotimes_{v\in\bar A} \ket{\Phi^+}_{u_vm_v}$. Consider any pure state $\ket{\varphi}_{\mathsf{A}E_A}$ and its Schmidt decomposition
    \begin{align}
        \ket{\varphi}_{\mathsf{A}E_A} = \sum_{j=1}^r \sqrt{\mu_j} \ket{x_j}_{\mathsf{A}} \ket{y_j}_{E_A},
    \end{align}
    where $\mu_j > 0$. Then,
    \begin{align}
        \ket{\varphi}_{\mathsf{A} E_A} \otimes \ket{\Phi}_T = \sum_{j=1}^r \sqrt{\mu_j} \ket{x_j}_{\mathsf{A}} \left( \ket{y_j}_{E_A} \otimes \ket{\Phi}_T \right).
    \end{align}
    The vectors $\ket{y_j}_{E_A} \otimes \ket{\Phi}_T,~j\in\{1,\ldots,r\}$, are orthogonal, and thus this is a Schmidt decomposition across $\mathsf{A}|E_A T$ with the same Schmidt coefficients. As a result,
    \begin{align}
        \mathrm{SR}_{\mathsf A|E_A T}
        \left(\ket{\varphi}_{\mathsf AE_A}\otimes\ket{\Phi}_T\right)
        =
        \mathrm{SR}_{\mathsf A|E_A}(\ket{\varphi}_{\mathsf AE_A}).
    \end{align}
    By the definition of Schmidt number, we have \eqref{eq:tensor}.
    
    Then, all switch operations act on the right side, and the processed users $A$ receive no new Bell links and no joint quantum operations across the bipartition. Hence, only classical communication crosses the fixed bipartition, and the remainder of the protocol is LOCC across the partition. Let $\Omega_{\mathrm{final}}$ denote the final joint state of all quantum systems. At this stage, let $\mathsf{B}$ denote all quantum systems held by the switch and the users in $\bar{A}$, including the remaining local ancillas.
    Since the target state must be distributed among the users, the final state factorizes as $\Omega_{\mathrm{final}} =\ketbra{\psi}_{A \bar{A}} \otimes \sigma_{{\mathrm{final}}}$, where $\sigma_{\mathrm{final}}$ is the joint state of all remaining systems. Recall that the Schmidt number of a pure state equals its Schmidt rank, and thus, 
    \begin{align}
        R_A = \mathrm{SN}_{A|\bar{A}} (\ketbra{\psi}_{A \bar{A}}) = \mathrm{SN}_{\mathsf{A}|\mathsf{B}}(\ketbra{\psi}_{A \bar{A}} \otimes \sigma_{\mathrm{final}}) \leq \mathrm{SN}_{\mathsf{A} | E_A T} \left( \omega_{\mathsf{A}E_A} \otimes \rho_T \right) \leq 2^{q_A}.
    \end{align}
    By taking the base-2 logarithm, we have \eqref{eq:general_memory_bound}. This completes the proof.
\end{Proof}

Furthermore, we now consider graph states. Since graph states are a class of stabilizer states, we first specify the lower bound in terms of the stabilizer group.

\subsection{Persistent Switch Memory Limit for Graph States}
We first consider the memory lower bound for stabilizer states. For a stabilizer state, the reduced state on the processed users $A$ is characterized by the stabilizers supported on those users. We define the \textit{reduced stabilizer group} as follows.
\begin{definition}
    For a stabilizer state $\ket{\psi}_{V}$ and $A \subseteq V$, the reduced stabilizer group on $A$ is defined as
    \begin{align}
        S_A
        \defeq
        \left\{
        s_A\in \mathcal{P}_A \mid  
        s_A \otimes I_{V\setminus A}\in S(\ket{\psi}_V)
        \right\},
    \end{align}
    where $\mathcal{P}_A$ denotes the Pauli group acting on the qubits in $A$, $S(\ket{\psi}_V)$ denotes the stabilizer group of $\ket{\psi}_V$. Under the embedding $s_A \mapsto s_A \otimes I_{V \setminus A}$, $S_A$ is a subgroup of the abelian stabilizer group $S(\ket{\psi}_V)$, and thus, it is also abelian. Let $d_A$ denote the number of independent generators of $S_A$, denoted by $s_1,\ldots,s_{d_A}$. Then, $|S_A| = 2^{d_A}$, where $|S_A|$ denotes the number of elements in the group $S_A$.
\end{definition}
Then, the reduced state on $A$ is determined by the stabilizers in $S_A$, shown as follows.
\begin{lemma}\label{lemma:S_A}
    The reduced state on $A$, i.e., $\rho_A \defeq \Tr_{V\setminus A} \ketbra{\psi}_V$, is described by
    \begin{align}
        \rho_A = \frac{1}{2^{|A|}} \sum_{s \in S_A} s = \frac{1}{2^{|A|}} \prod_{i=1}^{d_A} (I+s_i).
    \end{align}
\end{lemma}
\begin{Proof}
    Since $\ket{\psi}$ is a pure stabilizer state, we have \cite{gottesman2024surviving}
    \begin{align}
        \ketbra{\psi} = \frac{1}{2^n} \sum_{g \in S(\ket{\psi})} g.
    \end{align}
    Then,  
    \begin{align}
        \rho_A = \Tr_{\bar{A}} \ketbra{\psi} = \frac{1}{2^n} \sum_{g \in S(\ket{\psi})} \Tr_{\bar{A}} (g).
    \end{align}
    For every element $g \in S(\ket{\psi})$, we can write it as $g = g_A\otimes g_{\bar{A}}$, where $g_A$ and $g_{\bar{A}}$ act on $A$ and $\bar{A}$, respectively.
    Thus, 
    \begin{align}
        \Tr_{\bar{A}}(g) = g_A \Tr(g_{\bar{A}}).
    \end{align}
    Since $\Tr X = \Tr Y = \Tr Z = 0$ and $\Tr I = 2$, 
    if $g_{\bar{A}} \neq I_{\bar{A}}$, then $\Tr_{\bar{A}}(g) = 0$.
    This means that only stabilizers supported entirely on $A$ survive the partial trace.
    For such a stabilizer, 
    \begin{align}
        \Tr_{\bar{A}}(g_A \otimes I_{\bar{A}}) = g_A \Tr_{\bar{A}}(I_{\bar{A}}) = 2^{|\bar{A}|}g_A.
    \end{align}
    Hence,
    \begin{align}
        \rho_A = \frac{2^{|{\bar{A}}|}}{2^n} \sum_{s\in S_A}s = \frac{1}{2^{|A|}} \sum_{s\in S_A}s.
    \end{align}
    In addition, since the group $S_A$ is generated by $s_1,\ldots,s_{d_A}$, every element of $S_A$ has the form $s_1^{x_1} s_2^{x_2} \cdots s_{d_A}^{x_{d_A}}, x_i \in \{0,1\}$.
    Then, 
    \begin{align}
        \prod_{i=1}^{d_A}(I + s_i) = \sum_{x_i \in \mathbb{F}_2, i=1,\ldots,d_A } s_1^{x_1} \cdots s_{d_A}^{x_{d_A}}
        = \sum_{s \in S_A} s.
    \end{align}
    This completes the proof.
\end{Proof}

\begin{lemma}\label{lemma:rA-dA}
    Let $r_A \defeq |A| - d_A$. Then,  $\rank_{\mathbb{C}} (\rho_A) = 2^{r_A}$ and $S(\rho_A) = r_A$, where $S(\rho_A)$ is the von Neumann entropy of $\rho_A$.
\end{lemma}

\begin{Proof}
    For each stabilizer $s_i$, its $+1$ eigenspace projector is $(I+s_i)/2$. Hence, the projector onto the joint $+1$ eigenspace of $S_A$ is
    \begin{align}
        \Pi_A = \prod_{i=1}^{d_A} \frac{I+s_i}{2} = \frac{1}{2^{d_A}} \prod_{i=1}^{d_A} (I+s_i).
    \end{align}
    By Lemma~\ref{lemma:S_A}, we then have 
    \begin{align}
        \rho_A = 2^{-(|A|-d_A)} \Pi_A  = 2^{-r_A} \Pi_A. \label{reducedstate}
    \end{align}
    Since $\Pi_A$ is a projector, its rank equals its trace. Moreover,
    \begin{align}
        \Pi_A = \prod_{i=1}^{d_A} \frac{I+s_i}{2} = \frac{1}{2^{d_A}}\sum_{s \in S_A} s.
    \end{align}
    For the stabilizers in $S_A$, only the identity has a nonzero trace. Thus,
    \begin{align}
        \rank_{\mathbb{C}}(\Pi_A) = \Tr (\Pi_A) = \frac{1}{2^{d_A}} \Tr I_A =  2^{|A|-d_A} = 2^{r_A}. \label{projrank}
    \end{align}
    
    Since $\rho_A = 2^{-r_A} \Pi_A$, we have
    \begin{align}
        \rank_{\mathbb{C}}(\rho_A) = \rank_{\mathbb{C}}(\Pi_A) = 2^{r_A}.
    \end{align}
    Also, $\rho_A$ has exactly $2^{r_A}$ nonzero eigenvalues, all equal to $2^{-r_A}$, from \eqref{reducedstate} and \eqref{projrank}. Hence, the entropy of $\rho_A$ is 
    \begin{align}
        S(\rho_A) = -\Tr \rho_A \log_2 \rho_A = - \sum_{i=1}^{2^{r_A}} 2^{-r_A} \log_2 2^{-r_A} = r_A.
    \end{align}
    This completes the proof.
\end{Proof}
By Theorem~\ref{thm:memory_lower_bound}, any valid protocol therefore satisfies $q_A \geq r_A$ for a stabilizer target state. We next specialize this quantity to graph states.

\begin{lemma}\label{lemma:graph_state}
    Let $\ket{G}_V$ be the graph state associated with an undirected graph
    $G=(V,E)$, with binary adjacency matrix
    $\Gamma_G\in\mathbb{F}_2^{n\times n}$. The $(i,j)$th element of $\Gamma_G$ is defined by
    \begin{align}
        \Gamma_G(i,j) = \begin{cases}
            1, \quad \text{if}\ \{i,j\} \in E, \\
            0, \quad \text{otherwise}.
        \end{cases}
    \end{align}
    For any subset $A\subseteq V$, 
    \begin{align}
        S(\rho_A)=\rank_{\mathbb{F}_2}
    \Gamma_G[A,V\setminus A],\label{eq:S_graph}
    \end{align}
    where $\Gamma_G[A,V\setminus A]$ denotes the submatrix of $\Gamma_G$ whose rows and columns are indexed by $A$ and $V \setminus A$, respectively.
\end{lemma}
The identity in \eqref{eq:S_graph} is a standard result of the stabilizer representation of graph states \cite{hein2004multiparty}. For completeness, we provide a proof in Appendix~\ref{appendix:graph-state}.

Lemma~\ref{lemma:graph_state} describes $S(\rho_A)$ for any graph state. Based on this, we can characterize the switch-memory limit for graph-state distribution.

\begin{corollary}
    For distributing a graph state $\ket{G}_V$, after the Bell links of users in $A \subseteq V$ have been processed, the switch must retain at least
    \begin{align}
        r_A = \rank_{\mathbb{F}_2} \Gamma_G [A,V\setminus A] \label{eq:cut_rank}
    \end{align}
    persistent switch qubits.
\end{corollary}
Thus, for graph states, the cut-rank in \eqref{eq:cut_rank} gives a fundamental limit on the switch memory. In the next section, we achieve this limit by providing a computationally efficient stabilizer-transition protocol.

\section{Achieving the Memory Bound via Stabilizer Transitions}
Let $A\subseteq V$ denote the processed users and let $E$ denote the persistent qubits stored at the switch. We maintain a joint pure state $\ket\Psi_{AE}$ as a purification of the target reduced state $\rho_A \defeq \Tr_{V \setminus A} \ketbra{G}_V$. Our goal is to maintain a minimal stabilizer purification of $\rho_A$ using exactly $r_A$ persistent switch qubits after processing each user.

\subsection{Canonical Stabilizer Purification and Logical Mirrors}

\begin{definition}
A pure state $\ket{\Psi}_{AE}$ is a purification of $\rho_A$ if
\begin{align}
    \Tr_E \ketbra{\Psi}_{AE} = \rho_A.
\end{align}
It is a stabilizer purification if $\ket{\Psi}_{AE}$ is also a stabilizer state, and it is minimal if $|E|=r_A$.
\end{definition}
Since $\rank_{\mathbb{C}}(\rho_A)=2^{r_A}$, any purification of $\rho_A$ requires a purifier of dimension at least $2^{r_A}$. Hence, every purification uses at least $r_A$ switch qubits, and a minimal purification achieves this bound. 
In addition, a minimal purification need not be unique.
\begin{definition}
    Let $\ket{\Psi}_{AE}$ be a minimal stabilizer purification. For a Pauli operator $P_A \in N(S_A)$, a switch-side Pauli $\widetilde{P}_E$ is called a mirror of $P_A$ if
    \begin{align}
        P_A \otimes \widetilde{P}_E \in S(\ket{\Psi}_{AE}).
    \end{align}
\end{definition}

\begin{lemma}\label{lemma:unique_P_E}
    Let $\ket{\Psi}_{AE}$ be a minimal stabilizer purification for $\rho_A$. Then, ignoring global phases, every logical Pauli class in $N(S_A)/S_A$ has a unique switch-side Pauli mirror on $E$. 
    Moreover, the mirror map preserves the symplectic commutation relation.
\end{lemma}
\begin{Proof}
    First, note that every reduced stabilizer acts trivially on the switch side, i.e., $s_A \otimes I_E \in S(\ket{\Psi}_{AE}), \forall s_A \in S_A$. To see this, by the definition of $S_A$, we have
    \begin{align}
        (s_A\otimes I_{\bar A})\ket{G}_{V} = \ket{G}_{V}.
    \end{align}
    Note that the states $\ket{G}_V$ and $\ket{\Psi}_{AE}$ are both purifications of the same reduced state $\rho_A$, and $\ket{\Psi}_{AE}$ is a minimal purification. We then have
    \begin{align}
        \dim \mathcal{H}_E = \rank(\rho_A) \leq \dim \mathcal{H}_{\bar{A}}.
    \end{align}
    By the equivalence of purifications, there exists an isometry
    \begin{align}
        U: \mathcal{H}_E \rightarrow  \mathrm{supp}(\rho_{\bar A}), \qquad U^{\dagger} U = I_E
    \end{align}
    such that $\ket{G}_{V} = (I_A \otimes U) \ket{\Psi}_{AE}$.
    Therefore, 
    \begin{align}
        (I_A \otimes U) (s_A \otimes I_E) \ket{\Psi}_{AE} & = (s_A \otimes I_{\bar{A}}) (I_A \otimes U) \ket{\Psi}_{AE} \\
        & = (s_A \otimes I_{\bar{A}}) \ket{G}_{V} = \ket{G}_{V} = (I_A \otimes U) \ket{\Psi}_{AE}.
    \end{align}
    Applying $I_A \otimes U^{\dagger}$ to both sides and using $U^{\dagger} U = I_E$, we obtain
    \begin{align}
        (s_A\otimes I_E) \ket{\Psi}_{AE} = \ket{\Psi}_{AE}.
    \end{align}
    This implies that $s_A \otimes I_E \in S(\ket{\Psi}_{AE}), \forall s_A \in S_A$.
    
    Then, we show that there is no nontrivial stabilizer supported only on $E$, i.e., if $I_A \otimes P_E \in S(\ket{\Psi}_{AE})$, then $P_E = I_E$. To see this, since the purification is minimal, $E$ contains exactly $r_A$ qubits. Since $\ket{\Psi}_{AE}$ is pure, the entropies of $\rho_A$ and $\rho_E$ are the same, i.e., $S(\rho_A) = S(\rho_E) = r_A$. 
    The entropy of a state on a $d$-dimensional space is at most $\log_2 d$, with equality only for the maximally mixed state. Hence, $\rho_E = I_E/2^{r_A}$ and
    \begin{align}
        \rank_{\mathbb{C}} (\rho_E) = 2^{r_A} = \dim \mathcal{H}_E,
    \end{align}
    which implies that $\rho_E$ is full rank.
    Then, suppose $I_A \otimes P_E \in S(\ket{\Psi}_{AE})$,  then $P_E \rho_E = \rho_E$. Since $\rho_E$ is invertible, we have $P_E = I_E$.

    Now, we prove the existence of mirrors. Ignoring the global  phases, we define the map
    \begin{align}
        \phi: S(\ket{\Psi}_{AE}) \longrightarrow N(S_A)/S_A, \qquad \phi(P_A \otimes P_E) \defeq P_A S_A
    \end{align}
    This is a well-defined map, because for every element $P_A \otimes P_E \in S(\ket{\Psi}_{AE})$, it commutes with $s_A \otimes I_E, \forall s_A \in S_A$, hence, $[P_A,s_A] = 0$ and thus $P_A \in N(S_A)$, which means $P_A S_A$ is an element of $N(S_A)/S_A$. Moreover, $\phi$ is a group homomorphism. Specifically, for any $P_A \otimes P_E$, $Q_A \otimes Q_E \in S(\ket{\Psi}_{AE})$, we have
    \begin{align}
        \phi((P_A \otimes P_E) (Q_A \otimes Q_E)) = \phi(P_A Q_A \otimes P_E Q_E) = P_A Q_A S_A = (P_A S_A) (Q_A S_A).
    \end{align}
    Next, we consider the kernel of $\phi$. First, it can be seen that $S_A \otimes I_E \subseteq \mathrm{ker} \ \phi$. Conversely, suppose $P_A \otimes P_E \in \mathrm{ker} \ \phi$. Then, $P_A \in S_A$. Since $P_A \otimes I_E \in S(\ket{\Psi}_{AE})$, multiplying $P_A \otimes I_E $ and $P_A \otimes P_E$ gives $I_A\otimes P_E \in S(\ket{\Psi}_{AE})$. Since we have already proved that  $I_A \otimes P_E \in S(\ket{\Psi}_{AE}) \Rightarrow P_E = I_E$, we now obtain that
    \begin{align}
        \mathrm{ker} \ \phi = S_A \otimes I_E.
    \end{align}
    By the first isomorphism theorem, we have
    \begin{align}
        S(\ket{\Psi}_{AE}) /(S_A\otimes I_E) \cong \mathrm{im} \ \phi.
    \end{align}
    Since $\ket{\Psi}_{AE}$ is a pure stabilizer state on $|A| + r_A$ qubits, $|S(\ket{\Psi}_{AE})| = 2^{|A|+r_A}$, and since $|S_A| = 2^{d_A}$, we have
    \begin{align}
        |\mathrm{im} \  \phi|= \frac{|S(\ket{\Psi}_{AE})|}{|S_A|}= 2^{|A|+r_A-d_A} = 2^{2r_A},
    \end{align}
    where $d_A = |A| - r_A$. On the other hand, $S_A$ is a stabilizer group acting on $|A|$ qubits with $2r_A$ logical operators. Therefore, $|N(S_A)/S_A| = 4^{r_A}$. Since $\mathrm{im} \ \phi \subseteq N(S_A)/S_A$ and the two groups have the same cardinality, we have
    \begin{align}
        \mathrm{im} \ \phi  = N(S_A)/S_A,
    \end{align}
    i.e., the map $\phi$ is surjective. This means that for any logical Pauli class $P_A S_A \in N(S_A)/S_A$, there exists a stabilizer $Q_A \otimes \widetilde{P}_E \in S(\ket{\Psi}_{AE})$ such that $Q_A S_A = P_A S_A$. Therefore, there exists $s_A \in S_A$ such that $Q_A = P_A s_A$. Since $s_A \otimes I_E \in S(\ket{\Psi}_{AE})$, multiplying it with $P_A s_A \otimes \widetilde{P}_E$ gives $P_A \otimes \widetilde{P}_E \in S(\ket{\Psi}_{AE})$. Hence, every logical Pauli operator $P_A$ has a switch-side Pauli mirror.

    We next prove the uniqueness of $\widetilde{P}_E$. Suppose two representatives of the same logical class are $P_A$ and $P_A s_A, s_A \in S_A$, and suppose that their mirrors are $\widetilde{P}_E$ and $\widetilde{P}_E'$, respectively, i.e.,
    \begin{align}
        P_A \otimes \widetilde{P}_E \in S(\ket{\Psi}_{AE}), \qquad P_A s_A \otimes \widetilde{P}_E' \in S(\ket{\Psi}_{AE}).
    \end{align}
    By multiplying the second stabilizer with $s_A \otimes I_E$, we have $P_A\otimes \widetilde{P}_E' \in S(\ket{\Psi}_{AE})$. Thus,
    \begin{align}
        I_A \otimes \widetilde{P}_E\widetilde{P}_E' \in S(\ket{\Psi}_{AE}), 
    \end{align}
    which implies $\widetilde{P}_E = \widetilde{P}_E' $ up to some global phase. This proves the uniqueness.

    Finally, the mirror map also preserves multiplication. Specifically, suppose $P_A$ and $Q_A$ have mirrors $\widetilde{P}_E$ and $\widetilde{Q}_E$, respectively, i.e., $P_A \otimes \widetilde{P}_E \in S(\ket{\Psi}_{AE}), Q_A \otimes \widetilde{Q}_E \in S(\ket{\Psi}_{AE})$. Their product is also a stabilizer, which means
    \begin{align}
        P_A Q_A \otimes \widetilde{P}_E \widetilde{Q}_E \in S(\ket{\Psi}_{AE}).
    \end{align}
    Hence, $\widetilde{P}_E\widetilde{Q}_E$ is a mirror of $P_AQ_A$. Moreover, suppose that $P_A$ and $Q_A$ have the commuting relation as $P_A Q_A = (-1)^{c}Q_A P_A, c \in \{0,1\}$, and similarly, $\widetilde{P}_E \widetilde{Q}_E = (-1)^{c'} \widetilde{Q}_E \widetilde{P}_E, c' \in \{0,1\}$. Since $P_A\otimes\widetilde{P}_E$ and $Q_A\otimes\widetilde{Q}_E$ are both stabilizers of $\ket{\Psi}_{AE}$, they must commute. Therefore, their total commutation phase satisfies $(-1)^{c+c'} = 1$, which implies  $c=c'$ modulo $2$. Hence, the mirror map preserves commutation and anticommutation. This completes the proof.
\end{Proof}

We now introduce the canonical form for minimal stabilizer purifications.
\begin{definition}
    A minimal stabilizer purification is canonical if the generators of its stabilizer group are
    \begin{align}
        s_i\otimes I_E,
        &\qquad i=1,\ldots, d_A,
        \label{eq:canonical-old}\\
        \olX_{j}^A\otimes X_{E_j},\quad
        \olZ_j^A\otimes Z_{E_j},
        &\qquad j=1,\ldots,r_A,
        \label{eq:canonical-logical}
    \end{align}
    where $\{\olX_j^A,\olZ_j^A\}_{j=1}^{r_A}$ is a symplectic basis for $N(S_A)/S_A$, i.e., 
    \begin{align}
        \olX_i^A\olZ_j^A=(-1)^{\delta_{ij}}
        \olZ_j^A\olX_i^A,
    \end{align}
    modulo $S_A$. All $\olX_i^{A}$ commute with each other, and $\olZ_i^{A}$ commute with each other. These relations define a symplectic basis.
\end{definition}
In this form, each logical qubit of $\rho_A$ is paired with one switch qubit. Table~\ref{tab:canonical-purification} summarizes the canonical stabilizer purification.

It should be noted that the canonical form is not a restriction on the allowed minimal stabilizer purifications. In fact, every minimal stabilizer purification of $\rho_A$ can be transformed into the canonical form by a Clifford acting only on the switch-side qubits $E$. This allows us to use canonical stabilizer purifications as the input and output interfaces of the transition process.

\begin{theorem}\label{thm:canonical}
    Let $\ket{\Psi}_{AE}$ be any minimal stabilizer purification of $\rho_A$.
    For any logical symplectic basis
    $\{\olX_j^A,\olZ_j^A\}_{j=1}^{r_A}$, there exists a Clifford operator $C_E^{\mathrm{can}}$ acting only on $E$
    such that
    \begin{align}
        \ket{\Psi^{\mathrm{can}}}_{AE}
        =
        \left(
        I_A\otimes C_E^{\mathrm{can}}
        \right)
        \ket{\Psi}_{AE}
    \end{align}
    is a canonical stabilizer purification. Moreover, $C_E^{\mathrm{can}}$ can be implemented using $\CX$ and single-qubit Clifford gates.
\end{theorem}
\begin{Proof}
    By Lemma~\ref{lemma:unique_P_E}, each logical basis operator has a switch-side mirror. Let $\widetilde X_j^E$ and $\widetilde Z_j^E$ satisfy\footnote{The notation $\widetilde X_j^E$ and $\widetilde Z_j^E$ indicates that these operators mirror $\olX_j^A$ and $\olZ_j^A$, respectively. They may be arbitrary Pauli operators on $E$ and need not be purely $X$-type or $Z$-type.} 
    \begin{align}
        \olX_j^A\otimes\widetilde X_j^E
        \in
        S(\ket{\Psi}_{AE}),\qquad \olZ_j^A\otimes\widetilde Z_j^E
        \in
        S(\ket{\Psi}_{AE}).
    \end{align}
    Since the mirror map preserves the symplectic form, $ \left\{\widetilde X_j^E, \widetilde Z_j^E\right\}_{j=1}^{r_A}$ is a symplectic Pauli basis on the $r_A$ switch qubits.
    Furthermore, since any symplectic Pauli basis can be mapped to the standard Pauli basis by a Clifford operator \cite{dehaene2003clifford} using symplectic Gaussian elimination, there exists a switch-side Clifford $C_E^{\mathrm{can}}$ such that
    \begin{align}
        C_E^{\mathrm{can}}
        \widetilde X_j^E
        \left(C_E^{\mathrm{can}}\right)^\dagger = (-1)^{a_j}X_{E_j}, \quad C_E^{\mathrm{can}}
        \widetilde Z_j^E
        \left(C_E^{\mathrm{can}}\right)^\dagger
        =
        (-1)^{b_j}Z_{E_j}, \quad j=1,\ldots,r_A,
    \end{align}
    for some $a_j,b_j\in\{0,1\}$.
    The signs can be removed by an additional Pauli operator on $E$ \cite{dehaene2003clifford}. We also include these corrections in $C_E^{\mathrm{can}}$.
    
    Since $C_E^{\mathrm{can}}$ acts only on the switch, it does not change $s_A\otimes I_E$ and thus preserves the reduced state $\rho_A$.
    The resulting purification is therefore canonical. 
    Finally, note that the Clifford group is generated by the Hadamard transformation $H$, the phase gate $S$, and the $\CX$ gate \cite{gottesman2010introduction}, where
    \begin{align}
        H = \frac{1}{\sqrt{2}}\begin{bmatrix}
            1 & 1 \\
            1 & -1
        \end{bmatrix}, \qquad 
        S = \begin{bmatrix}
            1 & 0 \\
            0 & i
        \end{bmatrix}, \qquad
        \CX = \begin{bmatrix}
            1 & 0 & 0 & 0\\
            0 & 1 & 0 & 0\\
            0 & 0 & 0 & 1\\
            0 & 0 & 1 & 0
        \end{bmatrix}.
    \end{align}
    Then, $C_E^{\mathrm{can}}$ can be implemented using $H$, $S$, and $\CX$.
    This completes the proof.    
\end{Proof}

Consequently, every minimal stabilizer purification can be transformed to a canonical form.
\begin{table}[t]
    \centering
    \renewcommand{\arraystretch}{1.35}
    \[
    \begin{array}{cccc:ccc}
    &
    \multicolumn{3}{c}{\textbf{User side}}
    &
    \multicolumn{3}{c}{\textbf{Switch side}}
    \\
    &
    u_1 & \cdots & u_{|A|}
    &
    E_1 & \cdots & E_{r_A}
    \\
    \noalign{\global\arrayrulewidth=0.7pt}
    \cline{2-7}
    \noalign{\global\arrayrulewidth=0.4pt}
    \multirow{3}{*}{
    $S_A\!\left\{
    \vphantom{
    \begin{array}{c}
        s_1\\
        \vdots\\
        s_{d_A}
    \end{array}
    }
    \right.$
    }
    &
    \multicolumn{3}{c:}{s_1}
    &
    I & \cdots & I
    \\
    &
    \multicolumn{3}{c:}{\vdots}
    &
    \vdots & & \vdots
    \\
    &
    \multicolumn{3}{c:}{s_{d_A}}
    &
    I & \cdots & I
    \\
    \cdashline{2-7}
    &
    \multicolumn{3}{c:}{\olX_1^A}
    &
    X & \cdots & I
    \\
    &
    \multicolumn{3}{c:}{\olZ_1^A}
    &
    Z & \cdots & I
    \\
    &
    \multicolumn{3}{c:}{\vdots}
    &
    \vdots & \ddots & \vdots
    \\
    &
    \multicolumn{3}{c:}{\olX_{r_A}^A}
    &
    I & \cdots & X
    \\
    &
    \multicolumn{3}{c:}{\olZ_{r_A}^A}
    &
    I & \cdots & Z
    \end{array}
    \]
    \caption{Canonical minimal stabilizer purification across the user-side and the
    switch-side qubits.}
    \label{tab:canonical-purification}
\end{table}
Now, we are ready to present the complete stabilizer transition protocol. The construction follows an inductive invariant: after processing a new user's Bell link, the joint state of the processed users and the switch is a minimal stabilizer purification in canonical form. We first verify the invariant for the first processed user. Then, we show that each subsequent single-user transition preserves it.

\subsection{Initial Stage}
We first consider the arrival of the first user $v$. If $v$ is isolated in the target graph $G$, then its target state is $\ket{+}_{u_v}$. Thus, once the Bell pair link $\ket{\Phi^+}_{u_v m_v}$ is established, the switch measures $X_{m_v}$. For outcome $-1$, user $v$ applies $Z_{u_v}$. The switch then releases $m_v$, so the persistent memory is zero. We therefore assume below that $v$ has at least one neighbor in $G$.

Now, for a non-isolated $v$, the only stabilizer supported on $v$ alone is the identity. Hence, the reduced stabilizer group on $\{v\}$ is
\begin{align}
    S_{\{v\}} = \{I\},
\end{align}
which implies $d_{\{v\}} = 0, r_{\{v\}} = 1$.
Thus, the target reduced state of user $v$ is 
\begin{align}
    \rho_{\{v\}} = \frac{I_{u_v}}{2}.
\end{align}
Hence, when the Bell pair $\ket{\Phi^+}_{u_v m_v}$ is established, tracing out the switch-side half $m_v$ also gives
\begin{align}
    \Tr_{m_v} \ketbra{\Phi^+}_{u_v m_v} = \frac{I_{u_v}}{2} = \rho_{\{v\}}.
\end{align}
Therefore, the Bell pair is a purification of the target reduced state for the first processed user. In addition, it uses exactly $r_{ \{v\} } = 1$ switch qubit. Thus, this purification is a \textit{minimal} stabilizer purification. To see that it is canonical, the stabilizers are $X_{u_v} X_{m_v}$ and $Z_{u_v} Z_{m_v}$, and we can define
\begin{align}
    \olX_1^{\{v\}} = X_{u_v},
    \qquad
    \olZ_1^{\{v\}} = Z_{u_v}
\end{align}
such that the stabilizers are in the canonical form. Hence, the minimal canonical-purification invariant holds after the first processing step.

\subsection{Single-User Stabilizer Transitions}
Proceeding with the inductive step, suppose that the current state $\ket{\Psi}_{AE}$ is a minimal stabilizer purification in canonical form.

When the Bell link of the new user $v\notin A$ is generated, let $A' \defeq A \cup\{v\}$. The new link provides a Bell pair $\ket{\Phi^+}_{u_vm_v}$. Our goal is to transform $\ket{\Psi}_{AE} \otimes \ket{\Phi^+}_{u_v m_v}$ into a minimal purification for $A'$. To this end, we first show that the switch memory can change by at most one qubit per update.
\begin{lemma}\label{lemma_bounded_transition}
    For $A'=A\cup\{v\}$, we have
    \begin{align}
        |r_{A'} - r_A| \leq 1.
    \end{align}
\end{lemma}

\begin{Proof}
    For a graph state $\ket{G}$ with adjacency matrix $\Gamma_G$, $r_A = \rank(\Gamma_{G} [A,V\setminus A])$ and $r_{A'} = \rank \left(\Gamma_{G}[A\cup \{v\},V\setminus (A\cup \{v\})] \right)$.
    Let $c \defeq \rank(\Gamma_{G} [A,V\setminus (A\cup \{v\})])$.
    The old cut matrix adds one column to this common submatrix. The new cut matrix adds one row. Thus,
    \begin{align}
        r_A, r_{A'} \in \{c, c+1\},
    \end{align}
    which gives $|r_{A'} - r_A| \leq 1$. This completes the proof.
\end{Proof}
Lemma~\ref{lemma_bounded_transition} shows that every single-user update belongs to one of three cases:
\begin{align}
    r_{A'} - r_A \in \{+1,0,-1\}.
\end{align}
Recall that the number of independent generators in $S_A$ is $d_A=|A|-r_A$, and thus $d_{A'}=|A|+1-r_{A'}$. Then, $r_{A'} - r_A$ being $+1,0,-1$ corresponds to $d_{A'} - d_A$ being $0,1,2$, respectively, which are associated with the number of new generators that need to be added to $S_A$. The following lemma gives the precise form of these new generators. When comparing $S_A$ with $S_{A'}$, we use the natural embedding $s_A \mapsto s_A \otimes I_{u_v}$.

\begin{lemma}\label{lemma:structure}
    For $A' \defeq A \cup\{v\}$, the reduced stabilizer group $S_{A'}$ has the following structure.
    \begin{enumerate}
        \item If $r_{A'} = r_A + 1 $, then $S_{A'} = S_A \otimes I_{u_v}$, and thus $\rho_{A'} = \rho_A \otimes \frac{I_{u_v}}{2}$.

        \item If $r_{A'} = r_A$, then $S_{A'}$ contains exactly one new independent generator, which has the form
        \begin{align}
            g = P_A \otimes Q_{u_v}, \qquad Q_{u_v} \neq I,
        \end{align}
        where $P_A \in N(S_A)$. Hence, $S_{A'} = \langle S_A \otimes I_{u_v},g\rangle$.

        \item If $r_{A'} = r_A - 1$, then $S_{A'}$ contains exactly two new independent generators, which have the form
        \begin{align}
            g_X = P_A^X \otimes X_{u_v}, \qquad g_Z = P_A^Z \otimes Z_{u_v},\label{eq:g_XZ}
        \end{align}
        where $P_A^X,P_A^Z \in N(S_A)$, and $P_A^XP_A^Z = -P_A^Z P_A^X$.
        Hence, $P_A^X,P_A^Z$ form a logical Pauli pair for $S_A$, and
        $S_{A'} = \langle S_A \otimes I_{u_v},g_X,g_Z\rangle$.
    \end{enumerate}
\end{lemma}

\begin{Proof}
    We consider the quotient $S_{A'}/(S_A \otimes I_{u_v})$. Recall that $A' = A \cup \{v\}$, and thus $|A'| - |A|=1$. The quotient has dimension $\dim(S_{A'}/(S_A \otimes I_{u_v})) = d_{A'} - d_A$. Thus, this quotient describes exactly the independent stabilizer constraints that appear when the new user $v$ is added.

    Note that each element of $S_{A'}$ can be written as $P_A \otimes Q_{u_v}$, where $Q_{u_v}$ is a single-qubit Pauli operator. 
    We now recall the symplectic form of Pauli operators \cite{gottesman2024surviving}. Specifically, ignoring the phase, the single-qubit Pauli group can be represented by the binary vector space $\mathbb{F}_2^2$ through 
    \begin{align}
        I \leftrightarrow (0,0), \quad X \leftrightarrow (1,0), \quad Z \leftrightarrow (0,1), \quad Y \leftrightarrow (1,1). 
    \end{align}
    Now, ignoring global phases, consider any element in $S_{A'}$, denoted as $g = P_A \otimes Q_{u_v}$, where $Q_{u_v}$ corresponds to $ X_{u_v}^{x_v} Z_{u_v}^{z_v}, (x_v,z_v) \in \mathbb{F}_2^{2}$.
    We define the map
    \begin{align}
        \varphi_v: S_{A'}/(S_A \otimes I_{u_v}) \longrightarrow \mathbb{F}_2^2, \qquad \varphi_v(g S_A) \defeq (x_v,z_v).
    \end{align}
    Thus, $\varphi_v$ records only the Pauli operator acting on the newly added qubit $u_v$. This map is well-defined because two representatives of the same coset differ by an element $s_A \otimes I_{u_v}$, which does not change the Pauli acting on $u_v$. Moreover, the map is injective. To see this, suppose $\varphi_v(gS_A) = (0,0)$. Then, $g$ has the form $g = P_A \otimes I_{u_v}$. Since $g\in S_{A'}$ and $g$ is supported entirely on $A$, we have $P_A \in S_A$ by the definition of the reduced stabilizer group.
    Hence, $gS_A = S_A$, which is the zero element of the quotient $S_{A'}/(S_A \otimes I_{u_v})$. Thus, the kernel of $\varphi_v$ consists only of the zero coset, hence $\varphi_v$ is injective.

    We now consider the three possible rank changes. First, when $r_{A'} = r_A + 1$, then $d_{A'} - d_{A} = 0$. Hence, $S_{A'}$ and the embedded $S_A$ have the same dimension. Since $S_A \otimes I_{u_v} \subseteq S_{A'}$, we have $S_{A'} = S_A \otimes I_{u_v}$.
    Using the reduced-state stabilizer representation for the reduced state, we have
    \begin{align}
        \rho_{A'} = \frac{1}{2^{|A|+1}} \sum_{s_A\in S_A} s_A \otimes I_{u_v} = \left(
        \frac{1}{2^{|A|}}
        \sum_{s_A\in S_A}s_A \right) \otimes \frac{I_{u_v}}{2} = \rho_A \otimes \frac{I_{u_v}}{2}.
    \end{align}
    This proves the case of $r_{A'} = r_A + 1$.

    Second, when $r_{A'} = r_A$, we have $d_{A'} = d_{A} + 1$. Thus, $S_{A'}/S_A$ has dimension one. Therefore, exactly one new independent stabilizer generator is required. We can choose a representative of its nontrivial coset as $g = P_A \otimes Q_{u_v}$, where $Q_{u_v}$ cannot be the identity, because otherwise $g = P_A \otimes I_{u_v}$ would be supported entirely on $A$ and thus $g\in S_A$, contradicting $gS_A \neq S_A$. Furthermore, $g$ should commute with all embedded stabilizers $s_A \otimes I_{u_v}, s_A \in S_A$. This means $[P_A,s_A] = 0, \forall s_A \in S_A$, and thus $P_A \in N(S_A)$. This proves the case of $r_{A'} = r_{A}$.

    Finally, if $r_{A'} = r_A - 1$, we have $d_{A'} - d_A = 2$. This means that $\dim(S_{A'}/S_A) = 2$. Recall that the map $\varphi_v: S_{A'}/S_A \longrightarrow \mathbb{F}_2^2$ is injective. Since both its domain and codomain have dimension two, $\varphi_v$ is bijective. In particular, there exists a stabilizer coset whose restriction to $u_v$ has Pauli label $(1,0)$, corresponding to $X_{u_v}$, and another stabilizer coset that has the Pauli label $(0,1)$, corresponding to $Z_{u_v}$. Therefore, these two cosets can be characterized by two representatives $g_X$ and $g_Z$ defined in~\eqref{eq:g_XZ}. Since both $g_X$ and $g_Z$ commute with every embedded stabilizer $s_A \otimes I_{u_v}$, their $A$ parts $P_A^X, P_A^Z \in N(S_A)$. Moreover, since $g_X$ and $g_Z$ commute, and their $u_v$ parts $X_{u_v}, Z_{u_v}$ anticommute, their $A$ parts should anticommute, i.e., $P_A^XP_A^Z = - P_A^ZP_A^X $.
    Hence, $P_A^Z, P_A^X$ form an anticommuting logical Pauli pair of $N(S_A)/S_A$.
    This proves the case of $r_{A'} = r_A - 1$.
\end{Proof}

Lemma~\ref{lemma:structure} gives the exact stabilizer structure corresponding to each of the three rank changes. We next construct a switch-side operation that realizes each case while retaining exactly $r_{A'}$ persistent switch qubits after the transition.

\subsubsection{Rank-Increasing Transition}
When $r_{A'}=r_{A}+1$, we have $d_{A'} = d_A$, and therefore 
\begin{align}
    S_{A'} = \left\langle s_1 \otimes I_{u_v}, \ldots, s_{d_A} \otimes I_{u_v}\right \rangle.
\end{align}
The target reduced state $\rho_{A'}$ is $\rho_{A'} = \rho_A \otimes \frac{I_{u_v}}{2}$. Note that the current state
\begin{align}
    \ket{\Psi'}_{A' E m_v} \defeq \ket{\Psi}_{AE} \otimes \ket{\Phi^+}_{u_v m_v}
\end{align}
is already a stabilizer purification for $A'$. Also, we retain the new Bell half $m_v$ as a new persistent qubit at the switch, and the purification uses $r_A + 1 = r_{A'}$ persistent qubits. Hence, it is a minimal stabilizer purification.

Moreover, it is in the canonical form. Specifically, a logical symplectic basis for $S_{A'}$ can therefore be chosen as
\begin{align}
    \olX_j^{A'}
    &=
    \olX_j^A\otimes I_{u_v},
    \quad 
    \olZ_j^{A'}
    =
    \olZ_j^A\otimes I_{u_v},
    \qquad
    j=1,\ldots,r_A, 
    \\
    \olX_{r_A+1}^{A'}
    &=
    X_{u_v},
    \qquad \ 
    \olZ_{r_A+1}^{A'}
    =
    Z_{u_v}.
\end{align}
The resulting stabilizers are shown in Table~\ref{table:rank-increasing-purification}. Note that the old logical pairs remain mirrored by $E_1,\ldots,E_{r_A}$, while the stabilizers of the Bell pair, $X_{u_v}X_{m_v}$ and $Z_{u_v}Z_{m_v}$ are the new logical pair and corresponding mirror. Hence, the purification is canonical.

\begin{table}[t]
    \centering
    \renewcommand{\arraystretch}{1.35}
    \[
    \begin{array}{ccccc:cccc}
    &
    \multicolumn{4}{c}{\textbf{User side}}
    &
    \multicolumn{4}{c}{\textbf{Switch side}}
    \\
    &
    u_1 & \cdots & u_{|A|} & \blue{u_v}
    &
    E_1 & \cdots & E_{r_A} & \blue{m_v}
    \\
    \noalign{\global\arrayrulewidth=0.7pt}
    \cline{2-9}
    \noalign{\global\arrayrulewidth=0.4pt}
    \multirow{3}{*}{
    $S_{A'}\!\left\{
    \vphantom{
    \begin{array}{c}
    s_1\\
    \vdots\\
    s_{d_A}
    \end{array}
    }
    \right.$
    }
    &
    \multicolumn{3}{c}{s_1} & I
    &
    I & \cdots & I & I
    \\
    &
    \multicolumn{3}{c}{\vdots} & \vdots
    &
    \vdots & & \vdots & \vdots
    \\
    &
    \multicolumn{3}{c}{s_{d_A}} & I
    &
    I & \cdots & I & I
    \\
    \cdashline{2-9}
    &
    \multicolumn{3}{c}{\olX_1^A} & I
    &
    X & \cdots & I & I
    \\
    &
    \multicolumn{3}{c}{\olZ_1^A} & I
    &
    Z & \cdots & I & I
    \\
    &
    \multicolumn{3}{c}{\vdots} & \vdots
    &
    \vdots & \ddots & \vdots & \vdots
    \\
    &
    \multicolumn{3}{c}{\olX_{r_A}^A} & I
    &
    I & \cdots & X & I
    \\
    &
    \multicolumn{3}{c}{\olZ_{r_A}^A} & I
    &
    I & \cdots & Z & I
    \\
    \cdashline{2-9}
    &
    \multicolumn{3}{c}{I_A} & \blue{X}
    &
    I & \cdots & I & \blue{X}
    \\
    &
    \multicolumn{3}{c}{I_A} & \blue{Z}
    &
    I & \cdots & I & \blue{Z}
    \end{array}
    \]
    \caption{Canonical stabilizer purification after a rank-increasing transition. The original purification remains unchanged, while the fresh Bell pair contributes one additional logical pair and one new persistent switch qubit.}
    \label{table:rank-increasing-purification}
\end{table}

\subsubsection{Rank-Preserving Transition}\label{subsec:rank-preserving}
We now consider the case $r_{A'} = r_A$. Since $d_{A'}=d_A+1$, the reduced stabilizer group gains exactly one independent generator. Thus, we can write the reduced stabilizer group on $A'$ as
\begin{align}
    S_{A'} = \left\langle s_1 \otimes I_{u_v},\ldots,s_{d_A}\otimes I_{u_v}, g
    \right\rangle,
\end{align}
where the new generator is
\begin{align}
    g \defeq P_A \otimes Q_{u_v}, \quad Q_{u_v} \neq I, \quad P_A \in N(S_A).
\end{align}

By Lemma~\ref{lemma:unique_P_E}, we can choose a mirror $\widetilde{P}_E$  satisfying $P_A \otimes \widetilde{P}_E \in S(\ket{\Psi}_{AE})$. In a canonical stabilizer purification, $\widetilde{P}_E$ can be found from the logical expansion of $P_A$. Specifically, up to a global phase,
\begin{align}
    P_A = s_A \prod_{j=1}^{r_A} (\overline X_j^A)^{x_j} (\overline Z_j^A)^{z_j}, \qquad s_A\in S_A 
    \label{eq:PA-logical-expansion-physical}
\end{align}
for some $x_j, z_j \in \{0,1\}$. Then, its mirror $ \widetilde{P}_E$ has the symplectic form $(\mathbf{x} \mid \mathbf{z})$, where $\mathbf{x} = [x_1,\ldots,x_{r_A}]$, $\mathbf{z} = [z_1,\ldots,z_{r_A}]$. Then, $ \widetilde{P}_E$ is $ \prod_{j=1}^{r_A} X_{E_j}^{x_j} Z_{E_j}^{z_j}$.

Furthermore, the Bell pair satisfies
\begin{align}
    Q_{u_v} \otimes Q_{m_v}^T \in S(\ket{\Phi^+}_{u_v m_v}),
\end{align}
where $Q_{u_v} = Q_{m_v} \in \{X,Y,Z\}$. The transpose is needed because $(Q \otimes Q^T)\ket{\Phi^+} = \ket{\Phi^+}$. Now, let $M \defeq \widetilde{P}_E \otimes Q_{m_v}^T$, and then
\begin{align}
    g\otimes M \in S\left( \ket{\Psi}_{AE} \otimes \ket{\Phi^+}_{u_vm_v}
    \right)
\end{align}
is a stabilizer for the current state. Next, the switch measures $M$ with outcome $s$, which imposes the same eigenvalue $s$ on the user-side operator $g$. Then, the user applies a single-Pauli correction to make the eigenvalue $+1$. The detailed steps are as follows.

Before the switch measurement, the reduced state on the users $A'$ is
\begin{align}
    \rho_0 \defeq \rho_A \otimes \frac{I_{u_v}}{2}.
\end{align}
Let the measurement outcome of $M$ be $s \in \{+1,-1\}$. Since $g\otimes M$ stabilizes the current joint state, measuring $M$ with outcome $s$ projects the user subsystem onto the $s$-eigenspace of $g$. Let $\Pi_s \defeq (I+sg)/2$ denote the corresponding projector. The unnormalized post-measurement
user state is therefore
\begin{align}
    \widetilde{\rho}_{A'}^{(s)} = \Pi_s \rho_0 \Pi_s.
\end{align}
Since $g$ commutes with $\rho_0$ and $g^2=I$, this becomes
\begin{align}
    \widetilde{\rho}_{A'}^{(s)}=\frac{1}{2}(I+sg)\rho_0.
\end{align}
Moreover, since $g=P_A\otimes Q_{u_v}$ with $Q_{u_v}\neq I$, we have 
\begin{align}
    \Tr(g\rho_0)= \Tr(P_A\rho_A) \Tr\left(Q_{u_v}\frac{I_{u_v}}{2}\right) = 0,
\end{align}
where the last equality is because $\Tr(Q_{u_v}) = 0$. Hence, the probability of outcome $s$ is
\begin{align}
    p_s \defeq \Tr\widetilde{\rho}_{A'}^{(s)}=\frac{1}{2}.
\end{align}
Thus, the normalized post-measurement state is
\begin{align}
    \rho_{A'}^{(s)}=\frac{\widetilde{\rho}_{A'}^{(s)}}{p_s}=(I+sg)\rho_0.
\end{align}
Then, the target reduced stabilizer group is $S_{A'} = \langle S_A\otimes I_{u_v},g\rangle$, and the target reduced state is $\rho_{A'} = (I+g)\rho_0$. Thus, when $s = +1$, we have the target reduced state. When $s=-1$, we choose a single-qubit Pauli $R_{u_v}$ that anticommutes with $Q_{u_v}$, i.e.,
\begin{align}
    \{R_{u_v},Q_{u_v}\} = 0.
\end{align}
Applying $R_{u_v}$ flips the sign of the new generator, i.e.,
\begin{align}
    R_{u_v}(-g)R_{u_v}^{\dagger} = g.
\end{align}
Moreover, since all old generators act trivially on $u_v$, they are not changed by $R_{u_v}$. Hence, the resulting reduced stabilizer group is $S_{A'}$, and the reduced state on $A'$ is the target state $\rho_{A'}$.

Note that $M$ can act nontrivially on multiple switch qubits. To implement this measurement while releasing only $m_v$, let $\{e_1,\ldots,e_t\}$ denote the support of $M$ excluding $m_v$. Then, we can write the general form of $M$ as
\begin{align}
    M = P_{e_1} \cdots P_{e_t} Q_{m_v},
\end{align}
up to global phase, where each $P_{e_i}, i\in \{1,\ldots,t\}$, is not identity. For each Pauli, we apply a single-qubit Clifford mapping it to $Z$. Then, this operation maps $M$ as
\begin{align}
    M \mapsto  Z_{e_1} \cdots Z_{e_t} Z_{m_v},
\end{align}
up to a known sign.
Then, for each support qubit $e_i$, the switch applies $\CX_{e_i \rightarrow m_v}$, and then 
\begin{align}
    Z_{e_1} \cdots Z_{e_t} Z_{m_v} \mapsto Z_{m_v}.\label{eq:meas_Zmv}
\end{align}
Thus, $M$ is mapped to $ Z_{m_v}$. Hence, we have converted measuring $M$ to measuring $Z_{m_v}$.
     
After this measurement, $m_v$ is released. The remaining state is pure and stabilizer. Its reduced density matrix is $\rho_{A'}$, and exactly $r_A = r_{A'}$ switch qubits remain in the switch. Thus, it is a minimal stabilizer purification. Finally, the remaining purification can be canonicalized by a switch-side Clifford. This gives a minimal purification in canonical form. The resulting stabilizers are shown in Table~\ref{tab:transition-zero}.

\begin{table}[t]
    \centering
    \renewcommand{\arraystretch}{1.35}
    \[
    \begin{array}{ccccc:ccc}
    &
    \multicolumn{4}{c}{\textbf{User side}}
    &
    \multicolumn{3}{c}{\textbf{Switch side}}
    \\
    &
    u_1 & \cdots & u_{|A|} & u_v
    &
    E_1 & \cdots & E_{r_A}
    \\
    \noalign{\global\arrayrulewidth=0.7pt}
    \cline{2-8}
    \noalign{\global\arrayrulewidth=0.4pt}
    \multirow{4}{*}{
    $S_{A'}\!\left\{
    \vphantom{
    \begin{array}{c}
    s_1\\
    \vdots\\
    s_{d_A}\\
    P_AQ_{u_v}
    \end{array}
    }
    \right.$
    }
    &
    \multicolumn{3}{c}{s_1} & I
    &
    I & \cdots & I
    \\
    &
    \multicolumn{3}{c}{\vdots} & \vdots
    &
    \vdots & & \vdots
    \\
    &
    \multicolumn{3}{c}{s_{d_A}} & I
    &
    I & \cdots & I
    \\
    &
    \multicolumn{3}{c}{\blue{P_A}} & \blue{Q_{u_v}}
    &
    I & \cdots & I
    \\
    \cdashline{2-8}
    &
    \multicolumn{4}{c:}{\olX_1^{A'}}
    &
    X & \cdots & I
    \\
    &
    \multicolumn{4}{c:}{\olZ_1^{A'}}
    &
    Z & \cdots & I
    \\
    &
    \multicolumn{4}{c:}{\vdots}
    &
    \vdots & \ddots & \vdots
    \\
    &
    \multicolumn{4}{c:}{\olX_{r_A}^{A'}}
    &
    I & \cdots & X
    \\
    &
    \multicolumn{4}{c:}{\olZ_{r_A}^{A'}}
    &
    I & \cdots & Z
    \end{array}
    \]
    \caption{Canonical stabilizer purification after Transition~0. The new reduced stabilizer $P_A\otimes Q_{u_v}$ is incorporated into $S_{A'}$, while the switch retains $r_{A'}=r_A$ qubits.}
    \label{tab:transition-zero}
\end{table}

\subsubsection{Rank-Decreasing Transition}
We now consider the case $r_{A'} = r_A - 1$. In this case, we have $d_{A'} = d_A + 2$, which means that the reduced stabilizer group gains two independent generators. By Lemma~\ref{lemma:structure}, these two new generators can be chosen as
\begin{align}
    g_X = P_A^X \otimes X_{u_v}, \quad g_Z = P_A^Z \otimes Z_{u_v},
\end{align}
where both $P_A^X$ and $P_A^Z$ commute with $S_A$, while $P_A^X$ and $P_A^Z$ anticommute.
Thus, $P_A^X$ and $P_A^Z$ form a logical Pauli pair of the old reduced state $\rho_A$.

We first complete the logical operators to a symplectic basis
\begin{align}
    \left( P_A^X,P_A^Z \right),
    \left( \olX_2^A,\olZ_2^A \right),
    \ldots,
    \left( \olX_{r_A}^A,\olZ_{r_A}^A \right).
    \label{eq:minus-one-logical-basis}
\end{align}
This can be done by symplectic Gram-Schmidt orthogonalization \cite{wilde2009logical}.
By Lemma~\ref{lemma:unique_P_E}, we can find the mirrors corresponding to $P_A^X$ and $P_A^Z$ as $\widetilde{X}_{\rm out}^E$ and $\widetilde{Z}_{\rm out}^E$, respectively. Similarly, let $\left\{ \widetilde{X}_j^E,
\widetilde{Z}_j^E\right\}_{j=2}^{r_A}$ be the mirrors of the remaining logical pairs.

Now, we choose \textit{any} switch qubit $x \in E$ to be teleported to user $v$, and define the remaining switch register as $F \defeq E \setminus \{x\}$. By Theorem~\ref{thm:canonical}, there exists a switch-side Clifford $C_E$ such that
\begin{align}
    & \widetilde{X}_{\rm out}^E \mapsto X_x, \quad \widetilde{Z}_{\rm out}^E \mapsto Z_x, \\
    & \widetilde{X}_{j}^E \mapsto X_{f_{j}}, \quad \widetilde Z_j^E \mapsto Z_{f_j},
    \quad
    j=2,\ldots,r_A,
\end{align}
where $F = \{f_2,\ldots,f_{r_A}\}$.
After this Clifford, the purification is stabilized by
\begin{align}
    s_i\otimes I_E, & \qquad i=1,\ldots,d_A,
    \\
    P_A^X\otimes X_x, &\qquad P_A^Z\otimes Z_x,
    \\ \olX_j^A\otimes X_{f_j},
    &\qquad \olZ_j^A\otimes Z_{f_j}, \qquad j=2,\ldots,r_A.
    \label{eq:minus-one-decoded-purification}
\end{align}
Thus, the switch-side mirrors of the logical pair $(P_A^X,P_A^Z)$ are localized on the single qubit $x$. The stabilizers for the current state are shown in Table~\ref{tab:transition-minus-one-before-teleportation}. Then, by teleporting $x$ to $u_v$, the stabilizers $P_A^X \otimes X_{x} $ and $P_A^Z \otimes Z_{x} $ are transferred to $P_A^X \otimes X_{u_v} $ and $P_A^Z \otimes Z_{u_v} $, respectively.  

\begin{table}[t]
    \centering
    \renewcommand{\arraystretch}{1.35}
    \[
    \begin{array}{ccccc:ccccc}
    &
    \multicolumn{4}{c}{\textbf{User side}}
    &
    \multicolumn{5}{c}{\textbf{Switch side}}
    \\
    &
    u_1 & \cdots & u_{|A|} & \blue{u_v}
    &
    {\color{brown}{x}} & {f_2} & \cdots & {f_{r_A}} & \blue{m_v}
    \\
    \noalign{\global\arrayrulewidth=0.7pt}
    \cline{2-10}
    \noalign{\global\arrayrulewidth=0.4pt}
    \multirow{3}{*}{
    $S_A\!\left\{
    \vphantom{
    \begin{array}{c}
    s_1\\
    \vdots\\
    s_{d_A}
    \end{array}
    }
    \right.$
    }
    &
    \multicolumn{3}{c}{s_1} & I
    &
    I & I & \cdots & I & I
    \\
    &
    \multicolumn{3}{c}{\vdots} & \vdots
    &
    \vdots & \vdots & & \vdots & \vdots
    \\
    &
    \multicolumn{3}{c}{s_{d_A}} & I
    &
    I & I & \cdots & I & I
    \\
    \cdashline{2-10}
    &
    \multicolumn{3}{c}{\blue{P_A^X}} & I
    &
    {\color{brown}{X}} & I & \cdots & I & I
    \\
    &
    \multicolumn{3}{c}{\blue{P_A^Z}} & I
    &
    {\color{brown}{Z}} & I & \cdots & I & I
    \\
    \cdashline{2-10}
    &
    \multicolumn{3}{c}{\olX_2^A} & I
    &
    I & X & \cdots & I & I
    \\
    &
    \multicolumn{3}{c}{\olZ_2^A} & I
    &
    I & Z & \cdots & I & I
    \\
    &
    \multicolumn{3}{c}{\vdots} & \vdots
    &
    \vdots & \vdots & \ddots & \vdots & \vdots
    \\
    &
    \multicolumn{3}{c}{\olX_{r_A}^A} & I
    &
    I & I & \cdots & X & I
    \\
    &
    \multicolumn{3}{c}{\olZ_{r_A}^A} & I
    &
    I & I & \cdots & Z & I
    \\
    \cdashline{2-10}
    &
    \multicolumn{3}{c}{I_A} & \blue{X}
    &
    I & I & \cdots & I & \blue{X}
    \\
    &
    \multicolumn{3}{c}{I_A} & \blue{Z}
    &
    I & I & \cdots & I & \blue{Z}
    \end{array}
    \]
    \caption{Stabilizer representation before the teleportation step in Transition~$-1$. After the switch-side re-encoding, the logical pair $(P_A^X,P_A^Z)$ consumed by the new reduced stabilizers is localized on the switch qubit $E_x$, while the remaining $r_A-1$ logical pairs are mirrored by ${f_2},\ldots,{f_{r_A}}$. The fresh Bell pair $(u_v,m_v)$ contributes the stabilizers $X_{u_v}X_{m_v}$ and $Z_{u_v}Z_{m_v}$.}
    \label{tab:transition-minus-one-before-teleportation}
\end{table}

\begin{table}[t]
    \centering
    \renewcommand{\arraystretch}{1.35}
    \[
    \begin{array}{ccccc:ccc}
    &
    \multicolumn{4}{c}{\textbf{User side}}
    &
    \multicolumn{3}{c}{\textbf{Switch side}}
    \\
    &
    u_1 & \cdots & u_{|A|} & u_v
    &
    E_1 & \cdots & E_{r_A-1}
    \\
    \noalign{\global\arrayrulewidth=0.7pt}
    \cline{2-8}
    \noalign{\global\arrayrulewidth=0.4pt}
    \multirow{5}{*}{
    $S_{A'}\!\left\{
    \vphantom{
    \begin{array}{c}
    s_1\\[-1mm]
    \vdots\\[-1mm]
    s_{d_A}\\[-1mm]
    \olX_{r_A}^A X_{u_v}\\[-1mm]
    \olZ_{r_A}^A Z_{u_v}
    \end{array}
    }
    \right.$
    }
    &
    \multicolumn{3}{c}{s_1} & I
    &
    I & \cdots & I
    \\
    &
    \multicolumn{3}{c}{\vdots} & \vdots
    &
    \vdots & & \vdots
    \\
    &
    \multicolumn{3}{c}{s_{d_A}} & I
    &
    I & \cdots & I
    \\
    &
    \multicolumn{3}{c}{\blue{P_A^X}} & \blue{X}
    &
    I & \cdots & I
    \\
    &
    \multicolumn{3}{c}{\blue{P_A^Z}} & \blue{Z}
    &
    I & \cdots & I
    \\
    \cdashline{2-8}
    &
    \multicolumn{4}{c:}{\olX_1^{A'}}
    &
    X & \cdots & I
    \\
    &
    \multicolumn{4}{c:}{\olZ_1^{A'}}
    &
    Z & \cdots & I
    \\
    &
    \multicolumn{4}{c:}{\vdots}
    &
    \vdots & \ddots & \vdots
    \\
    &
    \multicolumn{4}{c:}{\olX_{r_A-1}^{A'}}
    &
    I & \cdots & X
    \\
    &
    \multicolumn{4}{c:}{\olZ_{r_A-1}^{A'}}
    &
    I & \cdots & Z
    \end{array}
    \]
    \caption{Canonical stabilizer purification after Transition~$-1$. A logical pair of the old purification is converted into two reduced stabilizers of $S_{A'}$ through a Bell measurement on $x$ and $m_v$. Consequently, both $x$ and $m_v$ are released, and the purifier size decreases from $r_A$ to $r_A-1$.}
    \label{tab:transition-minus-one}
\end{table}

To be specific, let $R \defeq AF$ denote all old systems except $x$. Before teleportation, the state can be written as
\begin{align}
    \ket{\phi}_{Rx} = \ket{\alpha}_R \ket{0}_x + \ket{\beta}_R \ket{1}_x. 
\end{align}
Together with the new Bell pair, the joint state is $\ket{\phi}_{Rx} \otimes \ket{\Phi^+}_{m_v u_v}$.
Applying $\CX_{x\rightarrow m_v}$ first and then $H_x$, the state becomes
\begin{align}
    H_x\CX_{x\rightarrow m_v}
    \left(\ket{\phi}_{Rx}\otimes\ket{\Phi^+}_{m_vu_v}
    \right) = \frac{1} {2}\sum_{a,b\in\{0,1\}}\ket{a}_x\ket{b}_{m_v}X_{u_v}^{b}Z_{u_v}^{a}\ket{\phi}_{Ru_v}.
    \label{eq:teleportation-identity}
\end{align}
Then, measuring $x$ and $m_v$ in the computational basis gives outcomes $a,b \in \{0,1\}$. 
The user $v$ then applies the Pauli correction $Z_{u_v}^{a} X_{u_v}^{b}$, which leads to the state $\ket{\phi}_{R{u_v}}$ up to a global phase.

Now, the subsystem carried by $x$ is transferred to $u_v$. The corresponding stabilizers are then transformed from $P_A^X \otimes X_x, P_A^Z \otimes Z_x$ to $P_A^X \otimes X_{u_v} = g_X, P_A^Z \otimes Z_{u_v} = g_Z$, respectively.
These are exactly the two new generators of $S_{A'}$.
The remaining logical operators remain mirrored by the qubits in $F$ and commute with both $g_X$ and $g_Z$. Note that $F$ has $r_A-1 = r_{A'}$ qubits. Hence, the new purification is minimal and canonical.

The resulting canonical stabilizer purification is shown in Table~\ref{tab:transition-minus-one}. Finally, the overall protocol is summarized in Algorithm~\ref{alg:stabilizer-transition}.

\begin{algorithm}[t]
\caption{Stabilizer-Transition Protocol Using Minimum Switch Memory}
\label{alg:stabilizer-transition}
\begin{algorithmic}[1]
\Require Target graph state $\ket{G}$.
\Ensure A minimal stabilizer purification in canonical form is maintained after each newly established Bell link is processed.

\State Initialize $A=\emptyset$, $E=\emptyset$, and $r_A=0$.

\For{each newly established Bell pair $\ket{\Phi^+}_{u_vm_v}$,}
    \State Set $A' = A\cup\{v\}$.
    \State Compute the reduced stabilizer groups $S_A$ and $S_{A'}$.
    \State Set $d_A$ and $d_{A'}$ as the number of independent generators of $S_A$ and $S_{A'}$, respectively.
    \State Compute
    $r_A=|A|-d_A$ and $r_{A'}=|A'|-d_{A'}$.

    \If{$r_{A'}=r_A+1$}
        \State Retain $m_v$ as a new switch qubit.
        \State Set $E \gets E \cup \{m_v\}$.

    \ElsIf{$r_{A'}=r_A$}
        \State Extract the new stabilizer
        $g=P_A\otimes Q_{u_v}$.
        \State Find the switch mirror $\widetilde P_E$ of $P_A$.
        \State Set
        $M=\widetilde P_E\otimes Q_{m_v}^{T}$.
        \State Reduce the measurement of $M$ to a signed $Z_{m_v}$ measurement and release $m_v$.
        \State Apply the Pauli correction on $u_v$.
        \State Canonicalize the new purification by a switch-side Clifford.

    \ElsIf{$r_{A'}=r_A-1$}
        \State Extract the two new stabilizers of $S_{A'}$.
        \State Identify the logical pair consumed by the two new stabilizers.
        \State Recode the consumed logical pair onto an output switch qubit $x\in E$.
        \State Re-encode the remaining logical pairs onto $E\setminus\{x\}$.
        \State Teleport $x$ to $u_v$ through the Bell pair $\ket{\Phi^+}_{m_vu_v}$.
        \State Release $x$ and $m_v$.
        \State Set $E \gets E \setminus \{x\}$ and $r_A \gets r_{A'}$.
    \EndIf

    \State Set $A\gets A'$.
\EndFor

\end{algorithmic}
\end{algorithm}

\begin{remark}
    If multiple Bell links succeed in the same round, the switch can process them sequentially in any internal order. Different orders can lead to different transient memory usage during the processing of the batch. However, in our model, the switch operations are assumed to be much faster than the Bell-link generation. Therefore, the persistent switch memory after processing the entire batch depends only on the set of processed users and is independent of the internal processing order.
\end{remark}
\begin{remark}
    Note that all steps in Algorithm~\ref{alg:stabilizer-transition} can be implemented in polynomial time using the binary symplectic representation of Pauli operators and stabilizer groups \cite{gottesman2024surviving,wilde2009logical,dehaene2003clifford}. In particular, bases of reduced stabilizer groups can be obtained by Gaussian elimination over $\mathbb{F}_2$, while logical Pauli bases can be constructed using symplectic Gram-Schmidt orthogonalization. The Pauli mirrors are obtained by solving the associated linear commutation constraints. Finally, the required switch-side Clifford operation is determined by the symplectic transformation between the two Pauli bases and can be implemented by symplectic Gaussian elimination. Since all matrices involved have dimension $O(n)$, each of these procedures can be implemented using at most $O(n^3)$ classical time in the worst case. Since there are $n$ such steps, Algorithm~\ref{alg:stabilizer-transition} has worst-case classical time complexity $O(n^4)$.
\end{remark}

\subsection{Forward Construction for Sequential Graph-State Generation}
We now show that the stabilizer-transition framework also provides a direct forward construction for sequential photonic graph-state generation \cite{emitter2022,schoen2005sequential}. Different from the quantum-switch setting where each remote user and the switch establish a Bell pair, sequential graph-state generation uses quantum emitters to generate entangled photons.
To be specific, consider a target graph state $\ket{G}_V$ associated with a connected graph $G=(V,E)$ with $n\geq 2$ vertices, where $V = \{1,\ldots,n\}$ denotes the indices of the photonic qubits. The photons are generated sequentially by quantum emitters, i.e., one at a time with an order $\pi=(v_1,\ldots,v_n)$. The goal is to eventually generate $\ket{G}_V$ using the minimum number of quantum emitters.

More specifically, consider the emission model of \cite{emitter2022}. All emitters are prepared in the initial $\ket{0}$ states. At each emission step, a fresh photon $p$ is initialized in $\ket{0}_p$ and can interact with an emitter $e$ through the emission operation $\CX_{e \rightarrow p}$, where the emitter $e$ is the control. We also allow joint Clifford operations among the emitters, single-qubit Clifford operations on the emitted new photon, and Pauli measurements on the emitters.

We now specify the protocol. For the emission order $\pi$, let $A_k \defeq \{v_1,\ldots,v_k\}$ denote the first $k$ emitted photons. $r_{A_k} \defeq \rank_{\mathbb{F}_2} (\Gamma_{G}[A_k,V\setminus A_k])$ is the number of active emitters in our protocol, which also is equal to the height function in \cite{emitter2022} at step $k$. Then, $r_{\max} \defeq \max_{0 \leq k \leq n} r_{A_k}$ is the number of emitters required throughout the process. We next show a forward construction with $r_{\max}$ emitters based on stabilizer-transition. 
The emitters are prepared in the product state $\ket{0}^{\otimes r_{\max}}$.
After the first $k$ photons have been emitted, let $E_k$ denote the active emitters. Similar to the quantum switch protocol, our construction maintains an invariant of canonical minimal stabilizer purification $\ket{\Psi_k}_{A_k E_k}$ of the target reduced state on $A_k$, such that
\begin{align}
    \Tr_{E_k} \ketbra{\Psi_k}{\Psi_k}_{A_k E_k} = \rho_{A_k}, \quad |E_k| = r_{A_k},\label{eq:emitter_invariant}
\end{align}
where $\rho_{A_k} \defeq \Tr_{V\setminus A_k} \ketbra{G}{G}_V$ is the target density operator on the first $k$ photons.
Our protocol then maintains this invariant by induction. To this end, we first describe the initial stage and then the single-photon transitions.

In the initial stage, consider the first photon $p = v_1$. Since $v_1$ is not isolated in the graph $G$, the reduced stabilizer group on $v_1$ is $S_{\{v_1\}} = \{I\}$. The number of active emitters then transforms from $r_{\emptyset} = 0$ to $r_{\{v_1\}} = 1$, and the target reduced state is $\rho_{\{v_1\}} = I_p/2$. Our operations are as follows. First, we apply $H_e$ on an emitter with state $\ket{0}_e$ to obtain $\ket{+}_e$. The photon is initialized as $\ket{0}_p$. We apply $\CX_{e \rightarrow p}$ and obtain the state $\ket{\Phi^+}_{ep}$. This produces the target reduced state since $\Tr_e\ketbra{\Phi^+}{\Phi^+}_{ep} = I_p/2$. Its stabilizer generators are $X_p X_e$ and $Z_p Z_e$, which are already in the canonical form. Thus, the invariant in \eqref{eq:emitter_invariant} holds after the first photon is emitted.

Now, suppose that after the first $k$ photons have been emitted, the state $\ket{\Psi_k}_{A_k E_k}$ satisfies the condition in \eqref{eq:emitter_invariant}. Let $p = v_{k+1}$ denote the next photon to be emitted, and $A_{k+1} = A_k \cup \{p\}$. By Lemmas~\ref{lemma_bounded_transition} and~\ref{lemma:structure}, we have $r_{A_{k+1}}-r_{A_k}\in\{-1,0,+1\}$.
Next, we use the stabilizer structure in Lemma~\ref{lemma:structure} to construct the emission operations in each case.

\noindent\emph{Case $1$: $r_{A_{k+1}}=r_{A_k}+1$.} In this case, we have the reduced stabilizer group on $A_{k+1}$ as $S_{A_{k+1}} = S_{A_k} \otimes I_p$, and the target reduced state is
\begin{align}
    \rho_{A_{k+1}} = \rho_{A_{k}} \otimes \frac{I_p}{2}.
\end{align}
We then activate one unused emitter $e$ and prepare it in $\ket{+}_e$ by applying $H_e$ to its initial state $\ket{0}_e$. The new photon $p$ is initialized as $\ket{0}_p$. We perform $\CX_{e \rightarrow p}$ to create $\ket{\Phi^+}_{ep}$. The resulting state is thus $\ket{\Psi_{k}}_{A_k E_k} \otimes \ket{\Phi^+}_{ep}$, which is a canonical minimal stabilizer purification of $\rho_{A_{k+1}}$ with $r_{A_{k+1}} = r_{A_k}+1$ active emitters.

\noindent\emph{Case $2$: $r_{A_{k+1}}=r_{A_k}$.} In this case, $S_{A_{k+1}}$ is obtained by adding one independent generator, i.e.,
\begin{align}
    S_{A_{k+1}} = \langle S_{A_k}\otimes I_p,g\rangle,
\end{align}
where
\begin{align}
    g \defeq P_A \otimes Q_p, \quad P_A \in N(S_{A_k}), \quad Q_p \neq I.
\end{align}
Note that $P_A$ represents a nontrivial logical Pauli class. Otherwise, the corresponding vertex of $p$ is isolated in the graph $G$.
By Lemma~\ref{lemma:unique_P_E}, we can find a mirror $\Tilde{P}_E$ of $P_A$, such that $P_A \otimes \Tilde{P}_E$ stabilizes the current state $\ket{\Psi_k}_{A_k E_k}$. Then, by emitter-side Clifford operations, we can map $\Tilde{P}_E$ to $Z_e$ on an active emitter $e$. After this, $P_A \otimes Z_e$ stabilizes the current state.
Now, the new photon $p$ has the initial state $\ket{0}_p$, stabilized by $Z_p$. We then apply the emission operation $\CX_{e \rightarrow p}$ such that the stabilizer changes as $Z_p \mapsto Z_e Z_p$. Then, $(P_A Z_e)(Z_e Z_p) = P_A Z_p$ stabilizes the current state. We can then apply a single-qubit Clifford $C_p$ to convert $Z_p$ to $Q_p$, which produces the desired stabilizer $g = P_A \otimes Q_p$. The resulting reduced stabilizer group is thus $S_{A_{k+1}}$, and the number of active emitters is $r_{A_{k+1}} = r_{A_k}$.

\noindent\emph{Case $3$: $r_{A_{k+1}}=r_{A_k}-1$.} By Lemma~\ref{lemma:structure}, there are two new independent stabilizers, which can be chosen as
\begin{align}
    g_X \defeq P_A^X \otimes X_p, \quad g_Z \defeq P_A^Z \otimes Z_p,
\end{align}
where $P_A^X, P_A^Z \in N(S_{A_k})$ and anticommute. Using an emitter-only Clifford in Theorem~\ref{thm:canonical}, we can map the mirrors of $P_A^X$ and $P_A^Z$ to $X_e$ and $Z_e$, respectively, on any active emitter $e$. The current state is then stabilized by $P_A^X X_e$ and $P_A^Z Z_e$. 
Now, the new photon $p$ has the state $\ket{0}_p$ and we apply $\CX_{e \rightarrow p}$ such that the stabilizers change as
\begin{align}
    P_A^X X_e \mapsto P_A^X X_e X_p, \quad P_A^Z Z_e \mapsto P_A^Z Z_e, \quad Z_p \mapsto Z_e Z_p.
\end{align}
By multiplying the last two stabilizers, we obtain $P_A^Z Z_p$. It remains to obtain $P_A^X X_p$. Since $P_A^X X_e X_p$ is a stabilizer of the current state, we can measure $X_e$. The outcome is denoted by $s \in \{+1,-1\}$, and the remaining system is then stabilized by $sP_A^X X_p$ and $P_A^Z Z_p$. If $s = -1$, then we apply $Z_p$ to obtain the stabilizer $P_A^X X_p$ because $Z_p X_p Z_p = -X_p$. After this, we obtain both $g_X$ and $g_Z$. The measured emitter $e$ is then disentangled and inactive. The remaining number of active emitters is then $r_{A_{k+1}} = r_{A_k} - 1$.

In each case, the output is a minimal stabilizer purification of $\rho_{A_{k+1}}$. By Theorem~\ref{thm:canonical}, we can apply emitter-side Clifford operations to produce a canonical form. Therefore, the inductive invariant \eqref{eq:emitter_invariant} is maintained. In the end, when $k=n$, we have $A_n = V$, $r_{A_n} = 0$. All emitters are inactive and decoupled, and the emitted photons are in the target graph state $\ket{G}_V$.

Recall that the number of active emitters is $r_{A_k}$ after the $k$th photon is emitted. Thus, the protocol requires $r_{\max} = \max_{0 \leq k \leq n} r_{A_k}$ emitters. This equals the maximum value of the height function in \cite{emitter2022}, which is the minimum number of emitters required for photonic-graph state generation.

\section{Illustrative Example}\label{sec:example}
To illustrate the proposed protocol, we consider a quantum switch network with four users. The users aim to build the four-qubit path graph $P_4: 1 - 2 -3 -4$ over their qubits $u_1,\ldots,u_4$. The stabilizer generators of $P_4$ are
\begin{align}
    K_1 &= X_{u_1}Z_{u_2},\quad 
    K_2 = Z_{u_1}X_{u_2}Z_{u_3}, \\
    K_3 &= Z_{u_2}X_{u_3}Z_{u_4},\quad  
    K_4 = Z_{u_3}X_{u_4}.
\end{align}
Use the order $\pi=(4,3,1,2)$, with $(4,3)$ as the internal order of the first batch in Figs.~\ref{fig:factory}--\ref{fig:factory_piecemaker}. The number of persistent switch qubits is $r_A: 0 \rightarrow 1 \rightarrow 1 \rightarrow 1 \rightarrow 0$. Hence, the switch needs at most one persistent qubit in the distribution process.

\begin{figure}[!t]
    \centering
    \includegraphics[width=1\textwidth]{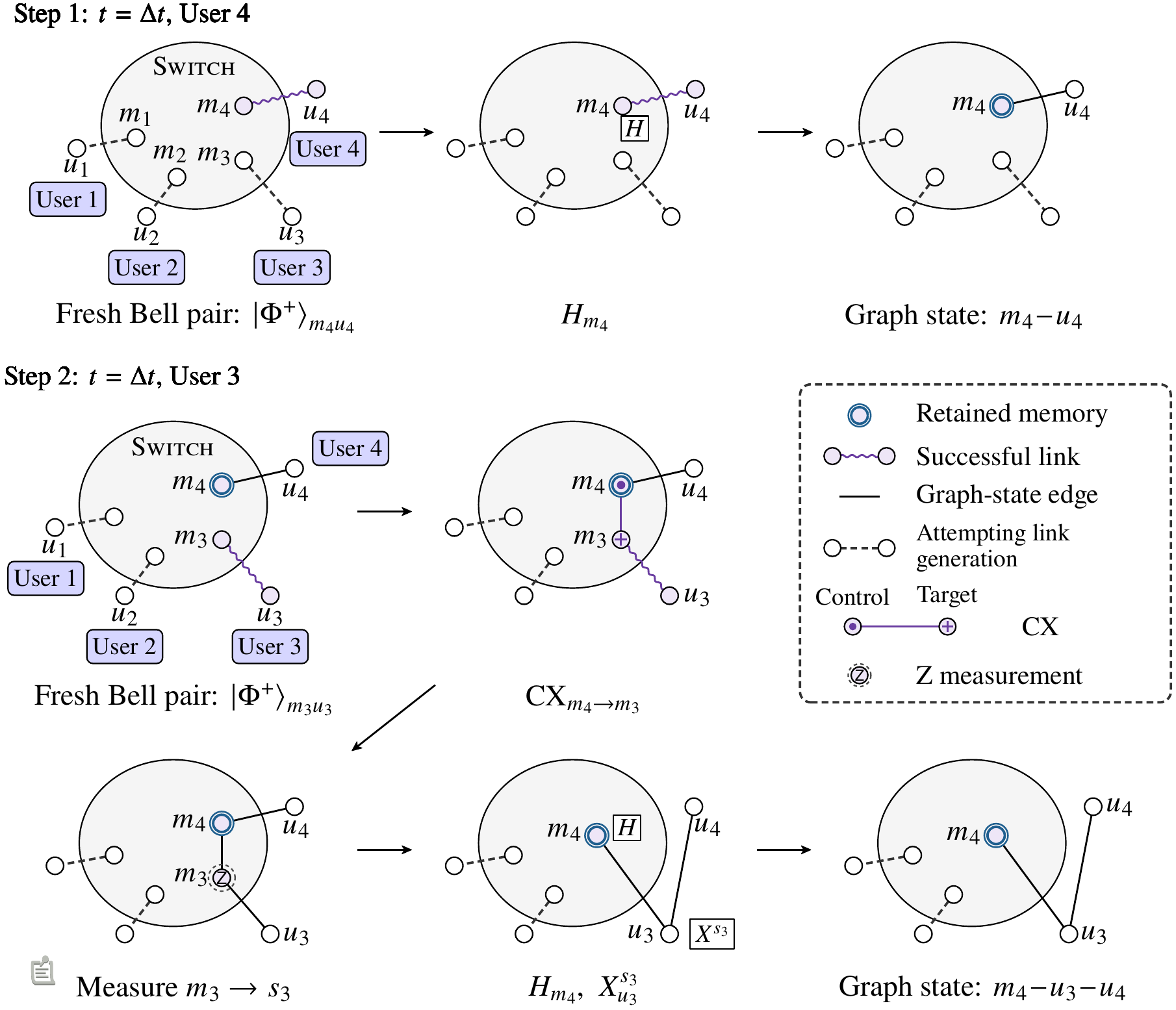}
    \caption{The first two steps of distributing the path graph $P_4$ under the Bell-link arrival order $\pi=(4,3,1,2)$. In Step $1$, the switch retains $m_4$ and applies $H_{m_4}$, which gives the graph state $m_4 - u_4$. In Step $2$, the switch processes the Bell pair for user $3$ using gates, measurement, and feed-forward correction shown in the figure. It releases $m_3$ and obtains the graph state $m_4-u_3-u_4$. Note that only $m_4$ remains at the switch after each step.}
    \label{fig:step_1_2}
\end{figure}
\paragraph{User 4: $+1$ transition.}
As shown in Step~1 of Fig.~\ref{fig:step_1_2}, the first successful link supplies the Bell pair $\ket{\Phi^+}_{u_4m_4}$ with stabilizers $X_{u_4}X_{m_4}$ and $Z_{u_4}Z_{m_4}$.
Since the required purifier size increases from $0$ to $1$, the switch retains $m_4$. For the convenience of illustrating the subsequent transitions, we apply $H_{m_4}$ to express the purification in a graph-state form.\footnote{This operation is not required by the proposed protocol. It is introduced only in this example to make the purification more transparent.} Thus, the stabilizers become $X_{u_4}Z_{m_4}$ and $Z_{u_4}X_{m_4}$,
which lead to the graph state
\begin{align}
    u_4-m_4.
\end{align}

\paragraph{User 3: $0$ transition.}
As shown in Step 2 of Fig.~\ref{fig:step_1_2}, the switch next processes the Bell pair $\ket{\Phi^+}_{m_3 u_3}$. For $A'=\{4,3\}$, the new reduced stabilizer is $g_3 = X_{u_4} Z_{u_3}$. Note that the current purification contains $X_{u_4} Z_{m_4}$, so the mirror of $X_{u_4}$ is $Z_{m_4}$. Together with the Bell pair for user $3$, this gives the switch-side Pauli measurement
\begin{align}
    M \defeq Z_{m_4} Z_{m_3}.
\end{align}
Then, we apply $\CX_{m_4\rightarrow m_3}$ which maps $M$ to $Z_{m_3}$, and then measure $m_3$ in the $Z$ basis. 

To explicitly track this transition, the joint purification and Bell pair are stabilized by 
\begin{align}
    X_{u_4} Z_{m_4}, \quad Z_{u_4} X_{m_4}, \quad X_{u_3} X_{m_3}, \quad Z_{u_3} Z_{m_3}.
\end{align}
After applying $\CX_{m_4\rightarrow m_3}$, these generators become
\begin{align}
    X_{u_4}Z_{m_4},\qquad Z_{u_4}X_{m_4}X_{m_3},\qquad X_{u_3}X_{m_3},\qquad Z_{u_3}Z_{m_4}Z_{m_3}.
\end{align}
Now, let $s_3 \in \{0,1\}$ denote the outcome of measuring $Z_{m_3}$. The corresponding eigenvalue is $s = (-1)^{s_3}$. Using the standard stabilizer measurement-update rule \cite{gottesman2024surviving}, we choose $X_{u_3}X_{m_3}$ as a generator that anticommutes with $Z_{m_3}$. Then, multiply the other generator that anticommutes with $Z_{m_3}$ by $X_{u_3}X_{m_3}$, and use the measured stabilizer $sZ_{m_3}$ to eliminate $Z_{m_3}$ from the remaining generators. We obtain the generating set after the measurement as
\begin{align}
    X_{u_4}Z_{m_4},\qquad Z_{u_4}X_{u_3}X_{m_4},\qquad sZ_{u_3}Z_{m_4},\qquad sZ_{m_3}.
\end{align}
The only remaining stabilizer acting on $m_3$ is $sZ_{m_3}$, so $m_3$ is disentangled from the other qubits and can be released. If $s=-1$, user $3$ applies $X_{u_3}$, which changes $sZ_{u_3}Z_{m_4}$ to $Z_{u_3}Z_{m_4}$. Thus, after the correction and release of $m_3$, the surviving purification can be stabilized by
\begin{align}
    X_{u_4}Z_{m_4},\qquad Z_{u_4}X_{u_3}X_{m_4},\qquad Z_{u_3}Z_{m_4}.
\end{align}
Replacing the first generator by its product with the third and then applying $H_{m_4}$ gives
\begin{align}
    X_{u_4}Z_{u_3}, \quad Z_{u_4}X_{u_3}Z_{m_4}, \quad Z_{u_3}X_{m_4}.
\end{align}
These are exactly the graph-state generators of
\begin{align}
    u_4-u_3-m_4.
\end{align}

\begin{figure}[!t]
    \centering
    \includegraphics[width=1\textwidth]{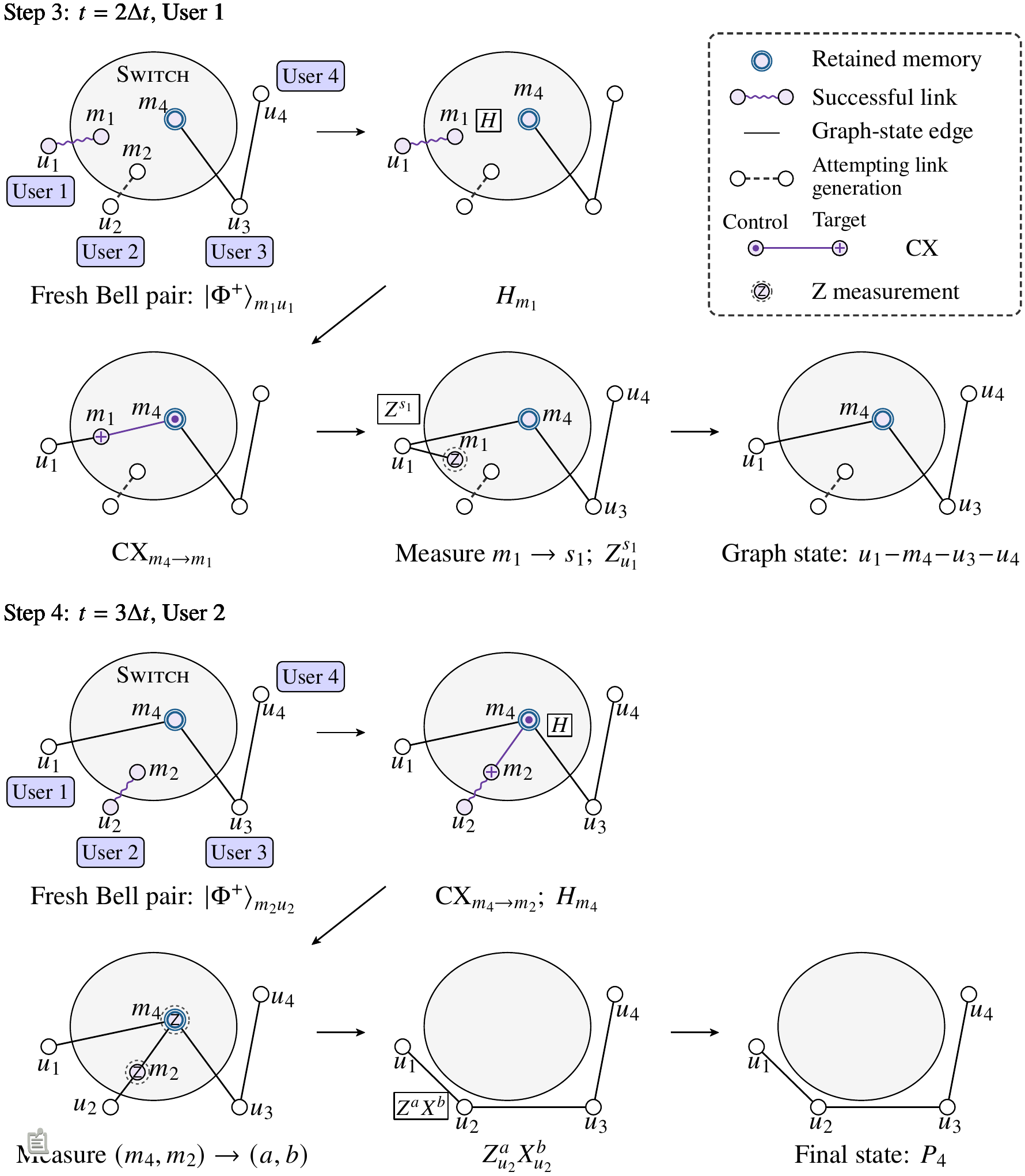}
    \caption{The final two steps of distributing the path graph $P_4$ under the Bell-link arrival order $\pi=(4,3,1,2)$. In Step $3$, the switch processes the new Bell pair for user $1$, using the operations shown in the figure. It releases $m_1$ and obtains the graph state $u_1-m_4-u_3-u_4$, retaining only $m_4$. In Step $4$, the switch teleports the state of $m_4$ to $u_2$ through the fresh Bell pair $\ket{\Phi^+}_{m_2u_2}$. After the Pauli correction, the users share the target graph state $u_1-u_2-u_3-u_4$. Both $m_4$ and $m_2$ are released, and no qubits remain stored in the switch.}
    \label{fig:step_3_4}
\end{figure}

\paragraph{User 1: $0$ transition.}
As shown in Step~3 of Fig.~\ref{fig:step_3_4}, the switch next incorporates user~1. For $A'=\{4,3,1\}$, the new reduced stabilizer is
\begin{align}
    g_1 = K_1K_3 = X_{u_1}X_{u_3}Z_{u_4}.
\end{align}
Since the current purification contains $Z_{u_4}X_{u_3}Z_{m_4}$, the mirror of $X_{u_3}Z_{u_4}$ is $Z_{m_4}$. Thus, let $M_1\defeq Z_{m_4}X_{m_1}$. By applying $H_{m_1}$ and then $\CX_{m_4 \rightarrow m_1}$, we map $M_1$ to $Z_{m_1}$. The switch measures $m_1$ in the computational basis and obtains the bit $s_1 \in \{0,1\}$. It then releases $m_1$. User $1$ applies $Z_{u_1}^{s_1}$, and the resulting purification is the graph state
\begin{align}
    u_4-u_3-m_4-u_1.
\end{align}
Now, the retained qubit $m_4$ has the same neighborhood in the purification graph as the target vertex $u_2$ has in $P_4$.

\paragraph{User $2$: $-1$ transition.}
As shown in Step 4 of Fig.~\ref{fig:step_3_4}, the switch processes the Bell pair between user $2$ and the switch. The required purifier size now decreases from $1$ to $0$. Thus, the reduced stabilizer group gains two independent generators. We choose
\begin{align}
    g_X &= K_2 = (Z_{u_1} Z_{u_3}) \otimes X_{u_2}, \\ 
    g_Z &= K_1 = X_{u_1} \otimes Z_{u_2}.
\end{align}
These generators have the form required by the general $-1$ transition, with
\begin{align}
    P_A^X = Z_{u_1} Z_{u_3}, \qquad P_A^Z = X_{u_1}.
\end{align}

Note that before user $2$ is processed, the purification is the graph state $u_4-u_3-m_4-u_1$. Its stabilizer group therefore contains
\begin{align}
     Z_{u_1}Z_{u_3}X_{m_4} = P_A^X\otimes X_{m_4}, \qquad X_{u_1}Z_{m_4} = P_A^Z\otimes Z_{m_4}. \label{eq:example-minus-logical-pair}
\end{align}
The mirrors of $P_A^X$ and $P_A^Z$ are already localized as $X_{m_4}$ and $Z_{m_4}$. Hence, the output qubit in the general protocol can be chosen as $x = m_4$, and no switch-side recoding is needed.

The switch then teleports $m_4$ to $u_2$ through the fresh Bell pair $\ket{\Phi^+}_{m_2 u_2}$. To be specific, the switch applies $\CX_{m_4 \rightarrow m_2}$ followed by $H_{m_4}$, and then measures $m_4$ and $m_2$ in the computational basis, obtaining the measurement outcomes $a,b \in \{0,1\}$, respectively. User $2$ applies $Z_{u_2}^{a} X_{u_2}^{b}$. After this correction, the stabilizers in \eqref{eq:example-minus-logical-pair} transform as
\begin{align}
    P_A^X\otimes X_{m_4} &\longmapsto P_A^X\otimes X_{u_2} = K_2, \\
    P_A^Z\otimes Z_{m_4} &\longmapsto P_A^Z\otimes Z_{u_2} = K_1.
\end{align}
Together with the surviving stabilizers $K_4$ and $K_1 K_3$, these generate the full stabilizer group of $P_4$. The final user state is thus the path state
\begin{align}
    u_1-u_2-u_3-u_4,
\end{align}
and both switch qubits, $m_4$ and $m_2$, are released. 

Overall, the protocol uses at most one persistent switch qubit throughout the entire distribution.

\section{Performance Evaluation}
We now compare the persistent switch-memory of the proposed protocol with that of Piecemaker protocol. We first show that for any graph and Bell-link arrival order, our protocol always requires less or equal persistent switch qubits than Piecemaker at any stage of the process. In particular, for the path graph $P_n$, our protocol reduces the peak switch memory from at least $\lfloor n/2\rfloor$ qubits in Piecemaker to one qubit. We then compare the cumulative switch-qubit storage time of the proposed protocol with Piecemaker and Factory protocols. We show that our protocol can significantly reduce the storage time, particularly for path graphs and tree graphs.

\subsection{Comparison of Switch Memory}
Piecemaker \cite{prielinger2026piecemaker} retains switch qubits corresponding to a vertex cover of a graph that is LC-equivalent to the target graph. To compare the number of switch qubits of the two protocols, we first establish the following bound.
\begin{lemma}\label{lemma:rank_VC}
    Let $G=(V,E)$ be a graph and let $C\subseteq V$ be any vertex cover of $G$. Then, for every $A\subseteq V$,
    \begin{align}
        \rank_{\mathbb F_2}
    \Gamma_G[A,V\setminus A] \leq |C|.\label{eq:cutrank_vertex_cover}
    \end{align}
\end{lemma}

\begin{Proof}
    Let $\bar{A} = V \setminus A$ and $\bar{C} = V \setminus C$.
    Note that
    \begin{align}
        \rank(\Gamma_G [A,\bar{A}])  = \rank\left(\begin{bmatrix}
            \Gamma_G [C\cap A, C\cap \bar{A}] & \Gamma_G[C\cap A, \bar{C}\cap \bar{A}] \\
            \Gamma_G[\bar{C}\cap A, C\cap \bar{A}] & \Gamma_G[\bar{C}\cap A, \bar{C}\cap \bar{A}]
        \end{bmatrix} \right),
    \end{align}
    where $C$ is a vertex cover and thus $\bar{C}$ is an independent set, i.e., 
    \begin{align}
        \Gamma_G[\bar{C}\cap A, \bar{C}\cap \bar{A}] = {0}.
    \end{align}
    Therefore, 
    \begin{align}
        \rank(\Gamma_G[A,\bar{A}])&\leq \rank\left(
        \begin{bmatrix} 
            \Gamma_G [C\cap A, C\cap \bar{A}] & \Gamma_G[C\cap A, \bar{C}\cap \bar{A}]] \end{bmatrix} \right) \nonumber \\
            &\quad+\rank\left(\begin{bmatrix}
                \Gamma_G[\bar{C}\cap A, C\cap \bar{A}] & \Gamma_G[\bar{C}\cap A, \bar{C}\cap \bar{A}]
        \end{bmatrix} \right)\\
            &\leq|C\cap A|+\rank\left(\begin{bmatrix}
                \Gamma_G[\bar{C}\cap A, C\cap \bar{A}] & 0]
            \end{bmatrix} \right)\\
            &\leq |C\cap A|+|C\cap \bar{A}| = |C|.
    \end{align}
    This completes the proof.
\end{Proof}

Piecemaker further exploits local-Clifford equivalence. Recall that a set $C$ is a local cover of the target graph $G$ if $C$ is a vertex cover of some graph $G'$ that is LC-equivalent to $G$. 
Since local operations preserve the Schmidt rank, we have
\begin{align}
    r_A = \rank_{\mathbb{F}_2} \Gamma_G [A, V \setminus A] = \rank_{\mathbb{F}_2}\Gamma_{G'} [A,V \setminus A].
\end{align}
Combining with the inequality in Lemma~\ref{lemma:rank_VC}, we have the following corollary.
\begin{corollary}\label{cor:r_A}
    Let $C$ be any local cover of $G$. Then,
    \begin{align}
        r_A = \rank_{\mathbb{F}_2} \Gamma_G[A,V\setminus A] \leq |C|,\quad \forall A \subseteq V.
    \end{align}
    In particular, this inequality holds for any minimal local cover selected by Piecemaker.
\end{corollary}

We now compare the memory usage of the proposed protocol with Piecemaker. Let $\pi \defeq (\pi_1,\ldots,\pi_n)$ denote the Bell-link arrival order, and let $A_k \defeq \{\pi_1,\ldots,\pi_k\} \subseteq V$ be the set of users whose Bell links have succeeded after the first $k$ successful link generations.
The proposed protocol processes each successful link and maintains a minimal stabilizer purification. After processing $A_k$, the switch retains
\begin{align}
    q^*(k) \defeq \rank_{\mathbb{F}_2} \Gamma_G[A_k,V \setminus A_k] 
\end{align}
qubits.

For Piecemaker, define the first local-cover arrival index
\begin{align}
    k_{\rm PM} \defeq \min\left\{
    k\in \{0,\ldots,n\} \mid \text{$A_k$ contains a local cover of $G$} \right\}.
\end{align}
At $k=k_{\rm PM}$, Piecemaker selects a minimal local cover $C\subseteq A_{k_{\rm PM}}$, where minimal means that no proper subset of $C$ is itself a local cover of $G$. By the definition of a local cover, there exists a graph $G'$ that is LC-equivalent to $G$ such that $C$ is a vertex cover of $G'$. Before this index, Piecemaker has not yet begun its stabilizer-processing stage, so every successful Bell-pair half remains stored
\begin{align}
    q_{\rm PM}(k)=k, \qquad k<k_{\rm PM}.
\end{align}
After the local cover becomes available, Bell-pair halves in $A_{k_{\rm PM}}\setminus C$ can be processed and released, and later arrivals outside $C$ can also be processed immediately. The qubits associated with $C$ remain persistently stored until the final step. Therefore, we have
\begin{align}
    q_{\rm PM}(k) = 
    \begin{cases}
        k,
        & k<k_{\rm PM},\\
        |C|,
        & k_{\rm PM}\le k<n,\\
        0,
        & k=n.
    \end{cases}
    \label{eq:pm-memory-profile}
\end{align}

\begin{theorem}
    For any $0 \leq k \leq n$, after processing the first $k$ arrivals, we have
    \begin{align}
        q^*(k) \leq q_{\rm{PM}}(k).
    \end{align}
\end{theorem}

\begin{Proof}
    For $k< k_{\mathrm{PM}}$, we have
    \begin{align}
        q^*(k) = \rank_{\mathbb F_2}\Gamma_G[A_k,V\setminus A_k] \leq |A_k| = k = q_{\rm{PM}}(k).
    \end{align}
    For $k_{\mathrm{PM}} \leq k < n$, Corollary~\ref{cor:r_A} directly gives $q^*(k) \leq |C| = q_{\mathrm{PM}}(k)$. At $k=n$, both protocols have released all switch qubits, and hence, $q^*(n) = q_{\mathrm{PM}}(n) = 0$. This completes the proof.
\end{Proof} 
Therefore, the proposed protocol always uses less or equal persistent switch memory than Piecemaker at any stage of the distribution process. 

Next, we use the path graph family to illustrate the memory gap between the two protocols. Consider the general path graph
\begin{align}
    P_n: 1-2-\cdots-n.
\end{align}
We have the following theorem showing that the MLC size of $P_n$, including all local-Clifford equivalences of $P_n$, is at least $\left\lfloor\frac{n}{2}\right\rfloor$.

\begin{theorem}\label{thm:path-local-cover}
    Let $\lambda_{\mathrm{LC}}(G)$ denote the minimum local cover size of $G$, i.e., 
    \begin{align}
        \lambda_{\mathrm{LC}}(G) \defeq \min_{G'\sim_{\mathrm{LC}}G} \tau (G'),
    \end{align}
    where $\tau (G')$ denotes the minimum vertex-cover size of $G'$. For the path graph $P_n$,
    \begin{align}
        \lambda_{\mathrm{LC}}(P_n) = \left\lfloor\frac{n}{2}\right\rfloor.
    \end{align}
\end{theorem}

\begin{Proof}
    The proof is trivial for $n=1$. Assume $n \geq 2$.
    Let $m \defeq \left\lfloor \frac{n}{2}\right\rfloor$, and partition the vertices of $P_n$ into the odd and even sets $O \defeq \{1,3,5,\ldots\} \cap V$, $D \defeq \{2,4,6,\ldots\} \cap V$, respectively. Then, $|D| = m$. Let $O' \defeq \{1,3,\ldots,2m-1\}$, and then the submatrix with rows $O'$ and columns $D$ is
    \begin{align}
        \Gamma_{P_n}[O',D] = \begin{bmatrix}
            1 & 0 & 0 & \cdots & 0 \\
            1 & 1 & 0 & \cdots & 0 \\
            0 & 1 & 1 & \ddots & \vdots \\
            \vdots & \ddots & \ddots & \ddots & 0 \\
            0 & \cdots & 0 & 1 & 1
        \end{bmatrix} \in \mathbb{F}_2^{m \times m}.
    \end{align}
    It has rank $m$. The full matrix $\Gamma_{P_n}[O,D]$ has only $m$ columns, so $\rank_{\mathbb{F}_2} \Gamma_{P_n}[O,D] = m$.
    Since cut-rank is invariant under local complementation, for every graph $G'$ that is LC-equivalent to $P_n$, we have $\rank_{\mathbb F_2} \Gamma_{G'}[O,D] = m$.
    Let $C$ be any vertex cover of $G'$. By Lemma~\ref{lemma:rank_VC}, we have $m \leq |C|$.
    Thus, every local cover of $P_n$ has at least $m$ vertices. The even set $D$ is a vertex cover of the original path and has size $ \left\lfloor n/2\right\rfloor$. Hence, $\lambda_{\mathrm{LC}}(P_n) = \left\lfloor n/2\right\rfloor$. This completes the proof.
\end{Proof}

\begin{remark}
    Theorem~\ref{thm:path-local-cover} shows that every local cover of $P_n$ contains at least $\left\lfloor\frac{n}{2}\right\rfloor$ vertices. Hence, Piecemaker cannot obtain a local cover before at least $\left\lfloor\frac{n}{2}\right\rfloor$ Bell links have succeeded. Before this occurs, all successful Bell-pair halves remain stored at the switch. Thus, in the whole distribution process of Piecemaker, the peak switch memory size $q_{\rm{PM}}^{\max}(P_n,\pi) \geq \left\lfloor\frac{n}{2}\right\rfloor$ for any Bell-link arrival order $\pi$.
\end{remark}

Next, we show that for a Bell-link arrival order $\pi = (1,2,\ldots,n)$, our protocol retains at most $1$ persistent qubit at the switch after each step.\footnote{The same result also holds for the reverse arrival order $\pi=(n,n-1,\ldots,1)$.}
Specifically, for any $1\leq k < n$, after the first $k$ links have been processed, the cut $A_k|V\setminus A_k$ contains only the edge $(k,k+1)$. Hence, 
\begin{align}
    q^*(k) = \rank_{\mathbb F_2} \Gamma_{P_n} [A_k,V \setminus A_k] = 1,\quad \forall 1\leq k < n,
\end{align}
and thus the peak memory size $q^*_{\max}(P_n,\pi)$ is $1$. For the same order $\pi$, the ratio of the two peak persistent switch memory satisfies
\begin{align}
    \frac{q^{\max}_{\rm{PM}}(P_n,\pi)}{q^*_{\max}(P_n,\pi)} \geq \left\lfloor\frac{n}{2}\right\rfloor. 
\end{align}
Thus, for this arrival order, the proposed protocol achieves a linear-factor reduction in peak switch memory compared with Piecemaker, reducing the memory scaling from $\Theta(n)$ to $\Theta(1)$.

\subsection{Cumulative Switch-Qubit Storage Time}
In the previous subsection, we compared the peak switch memory of Piecemaker and the proposed protocol. Now, to describe how long these qubits remain in the memory, we consider the \textit{cumulative switch-qubit storage time}. Specifically, let the Bell links succeed in the order $\pi = (\pi_1,\ldots,\pi_n)$ at times $t_1< \cdots < t_n$. For an entanglement distribution protocol, let $q(k)$ denote the number of switch qubits after the generation of the first $k$ Bell links. The cumulative switch-qubit storage time is defined as
\begin{align}
    B(G,\pi) \defeq \sum_{k=1}^{n-1} q(k) (t_{k+1} - t_{k}),
\end{align}
i.e., the sum of the storage times of all physical switch qubits. To explicitly compare the  protocols, we consider the case when exactly one Bell link succeeds at each round of $\Delta t$, i.e., $t_{k+1} - t_k = \Delta t$. The Factory protocol retains all successful Bell-pair halves until the last arrival. Thus, for \textit{any graph and arrival order}, the cumulative storage time of Factory is
\begin{align}
    B_{\mathrm{F}} = \sum_{k=1}^{n-1} k \Delta t = \frac{n(n-1)}{2} \Delta t = \Theta(n^2 \Delta t).
\end{align}

We first consider path graphs $P_n$. As shown in Theorem~\ref{thm:path-local-cover}, every local cover of $P_n$ has at least $m = \left\lfloor n/2 \right\rfloor$ vertices. In the Piecemaker protocol, the switch retains the first $k$ Bell-pair halves when $k < m$, and at least $m$ qubits for all $m \leq k < n$. Thus, we have $q_{\mathrm{PM}}(k) \geq \min \{k,m\}$ for every $1 \leq k \leq n-1$. Then, we have $\forall \pi$,
\begin{align}
    B_{\mathrm{PM}}(P_n,\pi) & \geq \sum_{k=1}^{n-1} \min \{k,m\} \Delta t  \\
    &=\left[\sum_{k=1}^{m-1}k+m(n-m)\right]\Delta t \\
    & = \begin{cases}
        \dfrac{n(3n-2)}{8}\Delta t,
        & n \text{ is even},\\
        \dfrac{(n-1)(3n-1)}{8}\Delta t,
        & n \text{ is odd}.
    \end{cases} 
\end{align}
Hence, $B_{\mathrm{PM}}(P_n,\pi) = \Omega(n^2 \Delta t)$. In addition, since $q_{\mathrm{PM}}(k)\leq k$, we also have $B_{\mathrm{PM}}(P_n,\pi)\leq B_{\mathrm F}=\Theta(n^2\Delta t)$. This implies that 
\begin{align}
    B_{\mathrm{PM}}(P_n,\pi)=\Theta(n^2\Delta t), \quad \forall \pi. \label{eq:cumulative-path-pm}
\end{align}
In contrast, under the order $\pi = (1,\ldots,n)$, our protocol requires $q^*(k) = 1, \forall 1 \leq k \leq n-1$. The cumulative storage time is then
\begin{align}
    B_{\mathrm{ST}}(P_n,\pi) = (n-1) \Delta t = \Theta(n \Delta t).
\end{align}
Hence, for this same arrival order, Piecemaker and Factory protocols require $\Theta(n^2 \Delta t)$ cumulative storage time, while the proposed protocol requires only $\Theta(n \Delta t)$.

Next, we extend the comparison to tree graphs. Let $T = (V,E)$ be a tree on $n \geq 2$ vertices, and let $d_{\max}$ denote its maximum degree. We have the following theorem to compare their cumulative storage times.
\begin{theorem}
    Consider any $n$-vertex tree graph $T$ with maximum degree $d_{\max}$. For any arrival order $\pi$, the cumulative storage time of Piecemaker satisfies
    \begin{align}
        B_{\mathrm{PM}}(T,\pi) \geq \frac{n(n-1)}{4 d_{\max}} \Delta t. \label{eq:cumulative-tree-pm}
    \end{align}
    In contrast, there exist some arrival orders $\pi'$ such that the cumulative storage time of the proposed protocol satisfies
    \begin{align}
        B_{\mathrm{ST}}(T,\pi') \leq (n-1)\lfloor\log_2 n\rfloor\Delta t. \label{eq:cumulative-tree-st}
    \end{align}
\end{theorem}

Recall that Factory protocol requires $B_{\rm F}=n(n-1)\Delta t/2$ for every arrival order. Hence, for tree graphs with constant $d_{\max}$, Piecemaker and Factory protocols require $\Theta(n^2\Delta t)$ cumulative storage time for every arrival order, whereas the proposed protocol achieves $O(n\log n\,\Delta t)$ for some arrival orders.

\begin{Proof}
    We first consider the Piecemaker protocol. Since every tree is bipartite, we can partition its vertex set as $V = L \cup R$, where $L$ and $R$ consist of vertices at even and odd distances from an arbitrary root of the tree, respectively. Then, each edge has one endpoint in $L$ and the other in $R$. Without loss of generality, we assume that $|L| \geq n/2$.

    Now, consider the matrix $\Gamma_T [L,R]$ and choose a set of linearly independent columns that form a basis of its column space. The number of selected columns is thus $\rank_{\mathbb{F}_2} (\Gamma_T [L,R])$. Since $T$ has no isolated vertices, each row of $\Gamma_T[L,R]$ is nonzero. Hence, every row must contain a nonzero entry in at least one basis column.
    On the other hand, since the tree $T$ has maximum degree $d_{\max}$, each column of $\Gamma_T [L,R]$ has at most $d_{\max}$ nonzero elements. Therefore, these basis columns can cover at most $d_{\max} \cdot \rank_{\mathbb{F}_2} (\Gamma_T [L,R])$ rows. 
    Hence, we have
    \begin{align}
        d_{\max} \cdot\rank_{\mathbb{F}_2} (\Gamma_T [L,R]) \geq |L| \geq \frac{n}{2}.
    \end{align}
    This gives $\rank_{\mathbb{F}_2} (\Gamma_T [L,R]) \geq n/(2d_{\max})$.
    Moreover, by Corollary~\ref{cor:r_A}, every local cover of $T$, denoted as $C$, has size at least $\rank_{\mathbb F_2}(\Gamma_T[L,R])$. Hence,
    \begin{align}
        |C|
        \geq \rank_{\mathbb F_2}(\Gamma_T[L,R])
        \geq \frac{n}{2d_{\max}}.
        \label{eq:bounded-tree-local-cover}
    \end{align}
    By the process of the Piecemaker protocol, the cumulative storage time is calculated as
    \begin{align}
        B_{\mathrm{PM}}(T,\pi)
        &\geq \sum_{k=1}^{n-1}
        \min\{k,|C|\}\Delta t \\
        &=\left[|C|(n-1)
        -\frac{|C|(|C|-1)}{2}
        \right]\Delta t \\
        &\geq \frac{|C|(n-1)}{2}\Delta t
        \\
        &\geq \frac{n(n-1)}{4d_{\max}}\Delta t.
    \end{align}
    This proves \eqref{eq:cumulative-tree-pm}.

    Next, we consider the proposed protocol. We construct an arrival order $\pi'$ recursively as follows. First, note that each tree $T$ has a centroid vertex $v$, whose removal splits the tree into connected components, denoted as $T_1,\ldots,T_s$. Each connected component contains at most $\lfloor n/2 \rfloor$ vertices \cite{durant2019centroid}. For each component $T_i$, we recursively construct an arrival order $\pi^{(i)}$ with the same procedure. We then define the order
    \begin{align}
        \pi' \defeq  (v,\pi^{(1)},\ldots,\pi^{(s)}),
    \end{align}
    where the vertices of component $T_i$ are in one block in $\pi'$.\footnote{Note that the construction of $\pi'$ is not unique. In fact, the orders $\pi^{(1)},\ldots,\pi^{(s)}$ can be arranged in any permutation and the inequality in \eqref{eq:tree-centroid-rank} still holds.} Let $\pi' = (v_1,v_2,\ldots,v_n)$ and $A_k \defeq \{v_1,\ldots,v_k\}$ denote the set of first $k$ arrived vertices. We then show by induction on $|V(T)|$ that 
    \begin{align}
        r_{\max}(T,\pi') \defeq \max_{0 \leq k \leq n}\mathrm{rank}_{\mathbb{F}_2} \Gamma_T[A_k,V\setminus A_k] \leq \lfloor\log_2 |V(T)|\rfloor,
        \label{eq:tree-centroid-rank}
    \end{align}
    where $|V(T)|$ denotes the number of vertices in $T$.

    First, the claim holds trivially for $|V(T)|=1$. For $|V(T)| = n \geq 2$, assume that \eqref{eq:tree-centroid-rank} holds for every tree with at most $n-1$ vertices. The cases of $k=0$ and $k=n$ are trivial. Now, suppose $1 \leq k \leq n-1$. Recall that $\pi' = (v,\pi^(1),\ldots,\pi^{(s)})$, where $\pi^{(i)}$ is the recursively constructed order for $T_i$.
    Since the component orders $\pi^{(1)},\ldots,\pi^{(s)}$ are consecutive blocks in $\pi'$, $A_k$, as a prefix, can cut at most one component, i.e., there exists at most one component $T_i$ such that $\emptyset \neq A_k \cap V(T_i) \neq V(T_i)$. If such a component $T_i$ exists, then $T_1,\ldots,T_{i-1}$ are contained in $A_k$ while $T_{i+1},\ldots,T_{s}$ are in $V\setminus A_k$.
    Moreover, since removing the centroid $v$ disconnects $T$ into the components $T_1,\ldots,T_s$, there are no edges between these components after removing $v$. Therefore, for the matrix removing the row corresponding to $v$, i.e., $\Gamma_T[A_k\setminus \{v\},V\setminus A_k]$, only the vertices of $T_i$ in $A_k$ and vertices of $T_i$ in $V \setminus A_k$ connect with each other. Hence, 
    \begin{align}
        \rank_{\mathbb F_2}
        (\Gamma_T[A_k\setminus \{v\},V\setminus A_k]) & \leq \rank_{\mathbb{F}_2} (\Gamma_T[A_k \cap V(T_i), V(T_i)\setminus (A_k \cap V(T_i))]).
    \end{align}
    Since $A_k\cap V(T_i)$ is a prefix of the order $\pi^{(i)}$ and $|V(T_i)| \leq\lfloor n/2\rfloor < n$, the induction hypothesis gives
    \begin{align}
        \rank_{\mathbb{F}_2} (\Gamma_T[A_k \cap V(T_i), V(T_i)\setminus (A_k \cap V(T_i))]) & \leq r_{\max} (T_i,\pi^{(i)}) \\
        & \leq \lfloor\log_2 |V(T_i)|\rfloor \leq \left\lfloor\log_2 \left\lfloor\frac{n}{2}\right\rfloor \right\rfloor.
    \end{align} 
    Since $v$ is the first vertex in $\pi'$, we have $v \in A_k$ for every $1 \leq k \leq n-1$. By separating the row corresponding to $v$ in $\Gamma_T[A_k,V\setminus A_k]$, we have
    \begin{align}
        \rank_{\mathbb F_2}
        (\Gamma_T[A_k,V\setminus A_k])
        \leq 1 + \rank_{\mathbb F_2}
        (\Gamma_T[A_k\setminus \{v\},V\setminus A_k]).
    \end{align}
    Combining the above bounds, we obtain
    \begin{align}
        \rank_{\mathbb F_2}
        (\Gamma_T[A_k,V\setminus A_k])
        &\leq 1+
        \left\lfloor\log_2 \left\lfloor\frac{n}{2}\right\rfloor \right\rfloor  \\
        &\leq \lfloor\log_2 n\rfloor.\label{eq:bound_log2n}
    \end{align}
    If no such component $T_i$ is partially intersected by $A_k$, then after removing the row corresponding to $v$, $\Gamma_T[A_k\setminus \{v\},V\setminus A_k]$ becomes a zero matrix. Hence, $\rank_{\mathbb F_2} (\Gamma_T[A_k,V\setminus A_k]) \leq 1$, and the same bound in \eqref{eq:bound_log2n} holds.
    This proves \eqref{eq:tree-centroid-rank}.

    Since our protocol retains $q^*(k) = \rank_{\mathbb F_2}  (\Gamma_T[A_k,V\setminus A_k])$ switch qubits after the first $k$ arrivals, we have
    \begin{align}
        B_{\mathrm{ST}}(T,\pi') & = \sum_{k=1}^{n-1} q^*(k) \Delta t \\
        & \leq (n-1) r_{\max}(T,\pi')\Delta t \\
        & \leq (n-1)\lfloor\log_2 n\rfloor \Delta t.
    \end{align}
    This proves \eqref{eq:cumulative-tree-st} and completes the proof.
\end{Proof}

\section{Conclusion}
In this paper, we characterized the minimum persistent quantum memory required for graph state distribution through a quantum switch. The lower bound of persistent switch memory applies to arbitrary pure states and reduces to cut-rank for graph states. We developed an online stabilizer-transition protocol that attains this bound. The construction requires no information about future Bell-link successes, and can be implemented in polynomial time.

We further proved that for every graph and every Bell-link arrival order, the proposed protocol uses less (or equal) persistent switch memory than Piecemaker through the distribution process. We also compared the cumulative switch-qubit storage time of the proposed protocol with Piecemaker and Factory protocols. Specifically, when exactly one Bell link succeeds in each round of duration $\Delta t$, for the path graph $P_n$, Piecemaker and Factory require $\Theta(n^2\Delta t)$ cumulative storage time for every arrival order, whereas our protocol requires only $(n-1)\Delta t$ under some orders. Furthermore, for bounded-degree trees, Piecemaker and Factory require $\Theta(n^2\Delta t)$ cumulative storage time for every arrival order, while our protocol achieves $O(n\log n\,\Delta t)$ under some arrival orders.

\appendix
\section*{Appendix}
\addcontentsline{toc}{section}{Appendix}

\section{Proof of Lemma~\ref{lemma:graph_state}}\label{appendix:graph-state}
\begin{Proof}
     By Lemma~\ref{lemma:rA-dA}, we have $S(\rho_A) = |A| - d_A$, where $d_A$ is the number of independent generators of $S_A$. We determine $d_A$ by identifying which products of graph-state generators are supported on $A$.
     
     Recall that for the target graph state $\ket{G}_V$, the $n$ independent generators of the stabilizer group are $K_v = X_v \prod_{w \in N_G(v)} Z_w, v \in V$.
     Every stabilizer therefore has a unique expression
     \begin{align}
         g(\mathbf{x})=\prod_{v\in V}K_v^{x_v},
        \qquad
        \mathbf{x}=(x_v)_{v\in V}\in\mathbb F_2^n.
     \end{align}
     Each generator contributes an $X$ on its own vertex and a $Z$ at each neighboring vertex. Thus, we can write
     \begin{align}
         g(\mathbf{x})= 
        \bigotimes_{u\in V}
        X_u^{x_u}
        Z_u^{\sum_{v\in V}\Gamma_G(u,v)x_v},
     \end{align}
     by ignoring the global phase. We now require $g(\mathbf{x})$ to act trivially on $\bar{A}$.
     For any $u \in \bar{A}$, if $x_u = 1$, its factor on $u$ is $X_u$ or $Y_u$, and cannot be the identity. Hence, we must have  $x_u = 0$ for every $u \in \bar{A}$. Then, every $s_A \in S_A$ has a unique representation as
    \begin{align}
        s_A \otimes I_{\bar{A}} = \prod_{v \in A} K_v^{x_v}, \qquad x_v \in \{0,1\}. \label{eq:x_v}
    \end{align}
    It remains to determine which choices of $\mathbf{x}_A \defeq (x_v)_{v \in A}$ give a product supported on $A$. We now fix a qubit $u \in \bar{A}$ and collect the contributions of all generators $K_v, v\in A$. For each $v\in A$, the factor of $K_v$ on $u$ is $Z_u^{\Gamma_G(u,v)}$. The product therefore acts on $u$ as
    \begin{align}
        \prod_{v\in A}Z_u^{\Gamma_G(u,v)x_v} = Z_u^{\sum_{v\in A}\Gamma_G(u,v)x_v}.
    \end{align}
    Since $Z_u^2=I_u$, this factor is the identity if and only if
    \begin{align}
        \sum_{v\in A}\Gamma_G(u,v)x_v=0,
    \end{align}
    over $\mathbb{F}_2$.
    Requiring this condition for all $u\in\bar A$ gives
    \begin{align}
        \Gamma_G[\bar{A},A]\mathbf{x}_A = \mathbf{0}.
    \end{align}
    Thus, the valid vectors $\mathbf{x}_A$ are precisely the elements of $ \mathrm{ker} \ \Gamma_G[\bar{A},A]$.
    By the uniqueness of the representation in \eqref{eq:x_v}, these vectors are in one-to-one correspondence with the elements of $S_A$. Therefore,
    \begin{align}
        d_A = \dim_{\mathbb{F}_2} (\mathrm{ker} \ \Gamma_G[\bar{A},A]) = |A| - \rank_{\mathbb{F}_2} \Gamma_G[\bar{A},A].
    \end{align}
    Moreover, since $G$ is undirected, we have $\Gamma_G = \Gamma_G^T$. Hence,
    \begin{align}
        d_A = |A| - \rank_{\mathbb{F}_2} \Gamma_G[{A},\bar{A}],
    \end{align}
    and finally,
    \begin{align}
        S(\rho_A) = |A| - d_A = \rank_{\mathbb{F}_2} \Gamma_G[{A},\bar{A}] = \rank_{\mathbb{F}_2} \Gamma_G[{A},V \setminus {A}].
    \end{align}
    This completes the proof.
\end{Proof}

\section*{Acknowledgement}
The authors acknowledge the assistance of ChatGPT 5.6 in this work. Specifically, it was used to identify relevant literature, improve the clarity and readability of the manuscript and to assist in preparing illustrative figures. Through a series of discussions, ChatGPT 5.6 also assisted in developing and checking proofs. In particular, it suggested generalizing the argument from Schmidt rank to Schmidt number in the proof of Theorem \ref{thm:memory_lower_bound} and proposed initial proofs of Lemma \ref{lemma:unique_P_E} and Lemma \ref{lemma:structure}. The authors developed and independently verified the final proofs and take full responsibility for the accuracy and content of the manuscript.

\bibliographystyle{unsrt}
\bibliography{Ref1.bib}
\end{document}